\documentclass[useAMS,usenatbib]{mn2e}
\usepackage{graphicx}

\def\solmas{$\mathrm{M_\odot}$\,}
\def\solmasp{$\mathrm{M_\odot}$}

\title[On column density thresholds and the star formation rate]{On column density thresholds and the star formation rate}
\author[Clark \& Glover]{Paul~C.~Clark$^{1, 2}$ \& Simon~C.~O.~Glover$^1$ \\
$^{1}$Zentrum f\"ur Astronomie der Universit\"at Heidelberg, Institut f\"ur Theoretische Astrophysik, Albert-Ueberle-Str.\ 2, 69120 Heidelberg \\
$^{2}$School of Physics and Astronomy, The Parade, Cardiff University, Cardiff, CF24 3AA \\
 {\tt email:} paul.clark@astro.cf.ac.uk, glover@uni-heidelberg.de}

\begin{document}

\maketitle

\begin{abstract}
We present the results of a numerical study designed to address the question of whether there is a column density threshold for star formation within molecular clouds. We have simulated a large number of different clouds, with volume and column densities spanning a wide range of different values, using a state-of-the-art model for the coupled chemical, thermal and dynamical evolution of the gas. We show that star formation is only possible in regions where the mean (area-averaged) column density exceeds $10^{21} \: {\rm cm^{-2}}$.  Within the clouds, we also show that there is a good correlation between the mass of gas above a K-band extinction $A_{\rm K} = 0.8$ and the star formation rate (SFR), in agreement with recent observational work. Previously, this relationship has been explained in terms of a correlation between the SFR and the mass in dense gas. However, we find that this correlation is weaker and more time-dependent than that between the SFR and the column density. In support of previous studies, we argue that dust shielding is the key process: the true correlation is one between the SFR and the mass in cold, well-shielded gas, and the latter correlates better with the column density than the volume density. 
\end{abstract}


\section{Introduction}
\label{sec:intro}
Ever since the pioneering work of \citet{schmidt59}, astrophysicists have been interested in 
understanding the relationship between the surface density of gas in a galaxy and the star
formation rate. Observational work by a large number of authors, summarised in the
comprehensive reviews of \citet{k98} and \citet{ke12} has demonstrated convincingly that on 
large scales ($\sim$~kpc or larger) there is a clear power-law relationship between the star 
formation rate surface density, $\Sigma_{\rm SFR}$, and the surface density of molecular 
hydrogen, $\Sigma_{\rm H_{2}}$, such that
\begin{equation}
\Sigma_{\rm SFR} \propto \Sigma_{\rm H_{2}}^{N}. \label{kslaw}
\end{equation}
Although the existence of this correlation -- commonly referred to as the Kennicutt-Schmidt law
-- is undisputed, its nature remains unclear. For instance, there is ongoing debate regarding the 
value of the power-law index $N$. Recent work by Bigiel and collaborators \citep{big08,big11,schr11} 
suggests that $N = 1.0 \pm 0.2$, consistent with a linear relationship. However, \citet{shetty13a,shetty13b} argue
that the data actually is more consistent with a sub-linear relationship, while other studies
argue for higher values of $N$, ranging from $N \sim 1.5$ \citep[e.g.][]{ken07,liu11} to $N \sim 2.0$ \citep[e.g.][]{nara12}.

Another important issue that remains uncertain is the extent to which this relationship is determined
by the physics of star formation within individual molecular clouds. If a correlation exists on the scale 
of individual clouds between their column densities and the rate at which they form stars, then one
would naturally expect to obtain a similar correlation on much larger scales, when one averages over 
many different clouds. However, this is not the only way to obtain such a correlation. Indeed, in the
extreme case in which the column densities of individual molecular clouds are uncorrelated with their 
star formation rates, then one can still obtain a large-scale, linear correlation between $\Sigma_{\rm SFR}$
and $\Sigma_{\rm H_{2}}$, provided only that there is a correlation between the number of clouds and
the number of star-formation regions, and that one is considering scales large enough to encompass 
many different examples of both, so that the small-scale stochasticity averages out. 

Observations of nearby molecular clouds provide some insight into this issue.
The fact that stars are observed to form only in dense molecular clouds, and not in diffuse molecular
or atomic clouds suggests that even if a relationship of the form of Equation~\ref{kslaw} holds on the
scale of individual clouds, it cannot hold for all clouds: there must be some minimum threshold
column density below which it breaks down. This hypothesis draws support from a number of simple
theoretical models that suggest that it should exist as a consequence of the fact that in order to form
stars efficiently, clouds must be able to shield themselves effectively from the external interstellar and 
extragalactic radiation fields \citep{schaye04,klm11,gc12a}. However, until now no detailed numerical
study of this idea has been carried out. 

In addition, several recent observational studies suggest that there is indeed a good correlation
between the local column density of gas and the star formation rate, but only within regions of
the cloud that have high column densities \citep{evans09,heid10,lada10,lada12}. These
studies show that star formation appears to occur only in gas with a total column density greater
than around $120 \: {\rm M_{\odot}} \: {\rm pc^{-2}}$, corresponding to an H$_{2}$ column density
of around $5 \times 10^{21} \: {\rm cm^{-2}}$, or a K-band extinction $A_{K} \simeq 0.8$. The
reason for this behaviour is currently unclear.

In light of this discussion, our aim in this paper is to address three key questions:
\begin{itemize}
\item Is there a mean column density below which clouds are unable to harbour star formation?
\item Within a star-forming cloud, does all star formation occur above some intrinsic column density, and if so, 
 then why does this occur?
\item Does the column density of a cloud affect the rate at which stars form within it, or
is this simply set by the mass volume density (and thus free-fall time) of the cloud \citep[see e.g.][]{km05,kt07}?
\end{itemize} 

Until very recently, it has been difficult to explore these issues using high resolution 3D simulations
of star formation within molecular clouds, as the models have lacked a sufficiently detailed and 
accurate model of the thermal physics of the clouds. However, recent developments 
\citep{gm07a,gm07b,g10,gc12a,gc12b,cgk12} have now made such a study possible. 

In this paper, we present results from a study in which we simulated a suite of clouds with different masses 
and sizes, selected so as to sample a wide range of different volume and column densities, including those 
found within regions of nearby star formation. Our simulations were performed using a modified version of
the {\sc GADGET-2} smoothed particle hydrodynamics (SPH) code \citep{springel05}. We have modified {\sc GADGET-2}
to include a detailed thermodynamical model of the gas that accounts for the main heating and cooling processes 
in the interstellar medium, capturing the line emission that dominates the cooling in the warm neutral medium
and cold neutral medium, as well as the dust cooling that occurs once the gas and dust are thermally coupled in 
pre-stellar cores.

The paper is arranged in the following manner. In Section \ref{sec:sims}, we present our suite of cloud simulations, along with details of our numerical model of the ISM and star formation. In Section \ref{sec:threshold} we discuss whether there exists a column density threshold, below which clouds are unable to form stars. The properties of the clouds, such as their temperature and density distributions are presented in Section \ref{sec:conditions}. We then show how the star formation rate (in our star-forming clouds) depends on the global properties of the cloud in Section \ref{sec:sfr}. In Section \ref{sec:ak08} we examine how the star formation rate correlates to the mass in ``dense'' gas, as discussed by Lada et al. (2010), and in Section \ref{sec:ak08-origin} we also suggest a new origin for this observed correlation, based on a simple shielding argument. The implications of this study are discussed in Section \ref{discuss} and we summarise the main findings of this study in Section \ref{conc}. 

\begin{figure}
\centerline{
\includegraphics[width=3.3in]{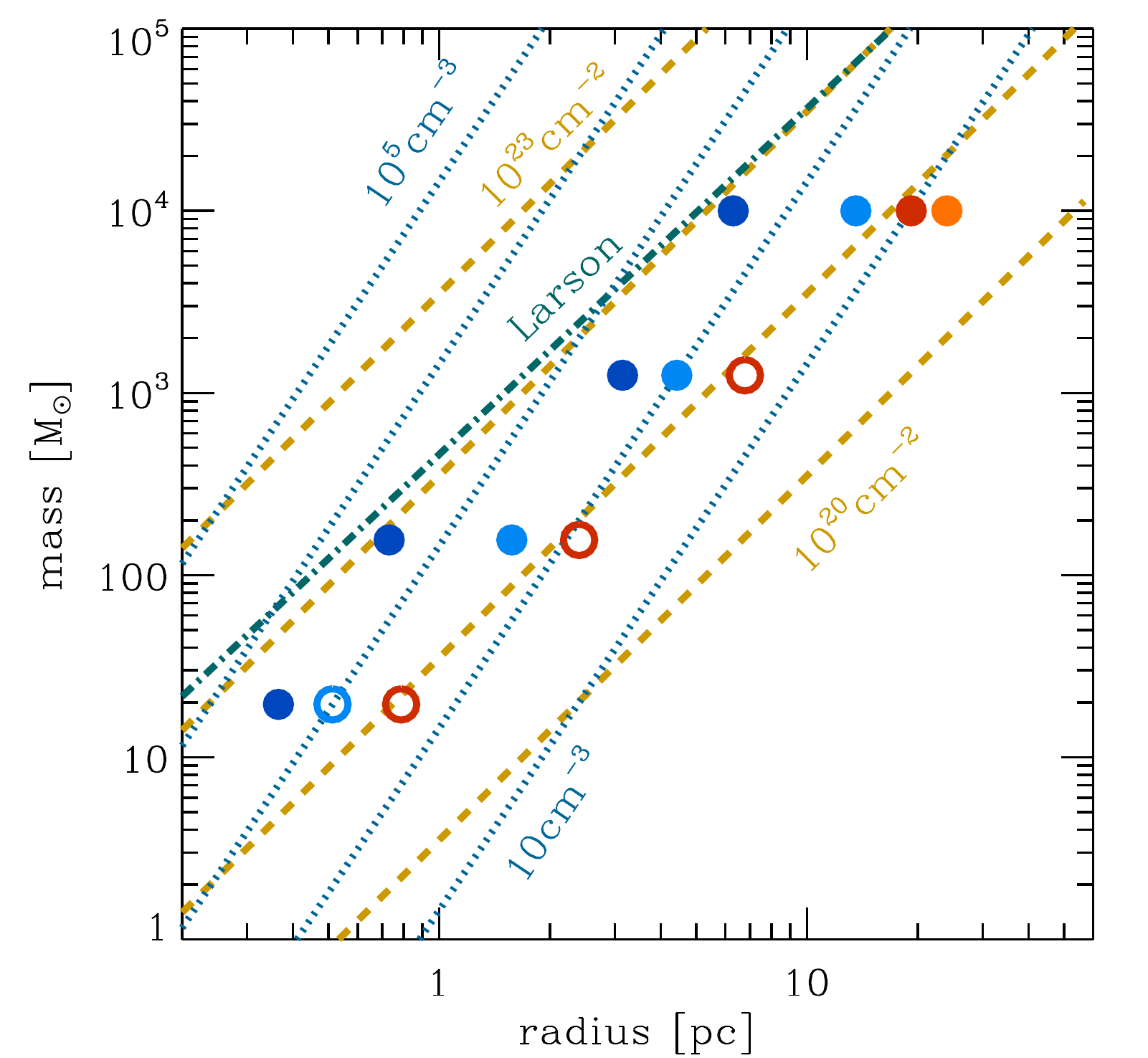}
}
\caption{Points denote the masses and radii of the clouds examined in this study. Clouds that were found  to be able to form stars are represented by filled circles, while clouds that were unable to form stars are represented by the open circles. For comparison, we plot both lines of constant column density (dotted) and volume density (dashed). The lines of constant column density were computed from $N = M/ \mu m_{\rm p} \pi r^2$, where $M$ and $r$ are the mass and radius of the cloud respectively, $\mu$ is the mean molecular weight and $m_{\rm p}$ is the proton mass. The lines of constant volume density were computed assuming a uniform sphere of mass $M$ and radius $r$. The dot-dashed line shows Larson's mass-radius relation \citep{larson81}. The numerical simulations are grouped by mass (20, 156, 1250, and 10,000 \solmasp). The colours of the points in this plot will be used to distinguish clouds of the same mass in subsequent plots, and are the same as those given in Table \ref{sim_table}. Note that we performed a second simulation with a mass of 10,000 \solmas and density of 264 cm$^{-3}$, which had a different turbulent seed. We have omitted this cloud from the figure for clarity, but it will appear in subsequent figures, and will be denoted by a green line/symbol.}
\label{mass-radius}
\end{figure}

\section{Simulations}
\label{sec:sims}

\begin{table*}
\caption{\label{sim_table} Overview of the clouds studied in this paper. The cloud identifiers, which will be used throughout the paper when discussing a particular model, are given by {\em n}`density'-{\em m}`mass', where `density' and `mass' are replaced by the number density and mass of the cloud in units of $\rm cm^{-3}$ and $\rm M_{\odot}$, respectively. For example, the first cloud listed in the table below has the identifer {\em n2640-m156}. Clouds are listed here first by decreasing initial number density ($n_0$), and then by decreasing mass. The initial column number densities ($N_0$) are calculated from $N = M/ \mu m_{\rm p} \pi r^2$, where $M$ and $r$ are the mass and radius of the cloud respectively, $\mu$ is the mean molecular weight and $m_{\rm p}$ is the proton mass.  All number densities quoted in this paper are with respect to the number of H nuclei. The turbulent velocities listed here are the full 3D RMS values. The SFR is calculated by averaging the combined accretion rate of all sink particles over the period from $t_{\rm SF}$, the time at which the first sink particle forms, to $t_{\rm end}$, the time at which the 10 percent of the cloud's initial mass has been accreted by the sink particles. More details can be found in Section \ref{sec:sfr}.}
\begin{center}
\begin{tabular}{c| l| l| l| c| c| c| c| c| c| c}
\hline
Mass & Symbol & Symbol & Line & Radius & $n_{0}$ & $N_{0}$ & v$_{\rm turb}$ & $t_{\rm SF}$ & $t_{\rm end}$ & SFR \\
$[\rm M_{\odot}]$ & shape & colour & style & [pc] & [cm$^{-3}$] & [cm$^{-2}$] & [km s$^{-1}$]& [Myr] & [Myr] & [\solmas yr$^{-1}$] \\

\hline \hline

156 & diamond &dark blue& dot-dashed& 0.7 & 2640 & $8.3 \times 10^{21}$ &  0.76	& 0.86 & 0.98 & $1.70\times10^{-4}$\\

20 & triangle &dark blue& dotted & 0.4 & 2640 & $4.1 \times 10^{21}$ &  0.36	& 1.08 & 1.16 &$1.94\times10^{-5}$\\

20 & triangle &light blue& dotted & 0.5 & 1000 & $2.1 \times 10^{21}$ & 0.32 & -- & 2.75 &--\\

10000 & square &dark blue& solid & 6.3 & 264	& $7.2 \times 10^{21}$ &  2.02	& 1.86 & 2.92 &$1.7\times10^{-3}$\\

10000 & square &green& solid & 6.3 & 264 & $7.2 \times 10^{21}$ &  2.02	& 1.85 & 2.52 &$2.64\times10^{-3}$\\

1250  & star &dark blue& dashed & 3.2 & 264	& $3.6 \times 10^{21}$ &  1.00	& 2.72 & 3.03 &$3.06\times10^{-4}$\\

156  & diamond &light blue& dot-dashed & 1.6 & 264 & $1.8 \times 10^{21}$ & 0.50 	& 3.55 & 3.70 &$1.12\times10^{-4}$\\

20 & triangle &red& dotted & 0.8 & 264 & $9.0 \times 10^{20}$ & 0.25 & -- & 5.35 &-- \\

1250 & star &light blue& dashed & 4.4 & 100& $1.8 \times 10^{21}$ &  0.86 & 4.62 & 4.98 &$1.21\times10^{-4}$\\  

156 & diamond &red& dot-dashed & 2.4  & 90	& $7.7 \times 10^{20}$ &  0.41	& -- & 9.17 &--\\

10000 &  square &light blue& solid & 13.6 & 26	& $1.5 \times 10^{21}$ &  1.4 & 8.97 & 10.3 &$7.44\times10^{-4}$\\

1250 & diamond &red& dashed  & 6.8  & 26	& $7.7 \times 10^{20}$ &  0.69	& -- & 17.07 &--\\

10000 & square &red& solid & 19.2 & 10	&$7.7 \times 10^{20}$ &  1.16	& 16.82 & 18.21 &$4.25\times10^{-4}$\\

10000 & square &orange& solid & 24.2 & 5 & $4.8\times 10^{20}$ & 1.03 & 29.78 & 37.79 & $1.18\times10^{-4}$ \\
\hline
\end{tabular}
\end{center}
\label{default}
\end{table*}

\subsection{The numerical model}
\label{sec:numerics}
We model the evolution of the gas in our simulations using a modified version of the {\sc GADGET-2} smoothed particle hydrodynamics code \citep{springel05}. We have modified the publicly available code in several respects. To begin with, we have added a sink particle implementation, based on the prescription in \citet{bbp95}, to allow us to follow the evolution of the gas beyond the point at which the first protostar forms. Our particular implementation is the one first described in \citet{jappsen05}. SPH particles are first considered as candidates for conversion to sink particles once their number density exceeds 10$^7$ cm$^{-3}$, the density at which the Jeans mass for 10~K gas becomes comparable to our mass resolution limit in the most poorly resolved simulation. However, sink particle creation only occurs if the candidate particle passes a number of tests:
the gas must be gravitationally bound, collapsing (negative velocity divergence), accelerating towards a common point (negative acceleration divergence), and be located further than a predefined distance from any other sink. In our simulations, this distance is set to 1200~AU, or roughly twice the expected Jeans length in gas with $n = 10^{7} \:
{\rm cm^{-3}}$ and $T = 10$~K.  Once a sink forms, it is free to accrete any gas that falls within the `accretion radius' of 600 AU, provided that it is not only bound to the sink 
particle, but is more tightly bound to that sink particle than to any other.

In addition, we have implemented an external pressure term \citep[e.g.][]{benz90} into the SPH equations. This enables us to model a constant pressure boundary, as opposed to the vacuum or periodic boundary conditions that are the only choices available in the standard version of {\sc GADGET-2}. The details of this term can be found in \citet{clark11}. To set the value of the external pressure term, $P_{\rm ext}$, we first run the code without it, and with the turbulent velocities set to zero, and let the outer layer of each cloud evolve to its equilibrium temperature.  The temperature and density of this outer layer are then used to calculate a value for $P_{\rm ext}$ for each cloud. Although we hold this value for the pressure constant during the course of the simulation, it should be noted that once the turbulence is included this pressure term is no longer strictly valid: the large-scale motions break the spherical symmetry in the cloud, and change the spatial variations in $A_{\rm V}$ that were responsible for setting the `initial' pressure. However, the pressure term is helpful for preventing the radial expansion that would naturally occur due to the pressure discontinuity when the simulation is started.

To model the chemical and thermal evolution of the gas, we use the treatment described in \citet{gc12b,gc12c}. This treatment combines the network for hydrogen chemistry
introduced by \citet{gm07a,gm07b} with an approximate model of CO formation and destruction proposed by \citet{nl99}. Comparison of this simple CO model against the
more complex model presented in \citet{g10} demonstrates that it produces comparable results at roughly a third of the computational cost \citep{gc12b}. Our model includes
H$_{2}$ formation on dust grains (following the classic prescription of \citealt{hm79}), but does not include any other grain surface processes, such as the freeze-out of CO.
However, we do not expect this omission to significantly affect the thermal balance of the gas \citep{gold01}.
 
We assume that our clouds are illuminated by the standard interstellar radiation field (ISRF), using the parameterisation of \citet{dr78} for the ultraviolet portion of the spectrum,
and that of \citet{bl94} at longer wavelengths. We keep the radiation field strength fixed in all of our models. To treat the penetration of radiation into the model clouds, we use the {\sc TreeCol} algorithm \citep{cgk12}, as described in \citet{gc12a}. Our treatment accounts for the effects of dust extinction, H$_{2}$ self-shielding (following \citealt{db96}), CO self-shielding and the shielding of CO by H$_{2}$ \citep{lee96}. 
We take the cosmic ray ionization rate (CRIR) to be $\zeta_{H} = 3 \times 10^{-17} \: {\rm s^{-1}}$. This is somewhat lower than the values now thought to be appropriate for the
diffuse ISM \citep[see e.g.][]{mac03,shaw08,ind12}, but is consistent with the value determined for gas within dense prestellar cores \citep[see e.g.][]{vdt00,mb07}. 
Variations in this value of a factor of a few are unlikely to significantly affect our results \citep{gc12c}.

In addition to our 3D models, whose initial conditions are described in section~\ref{sec:clouds} below, 
we also have run a large suite of one-zone models using the same chemical and 
thermal treatment as in the dynamical models. For each of these one-zone models, we specify the gas number density $n$ and the total column density $N$. The latter 
quantity is then used to compute the attenuation of the ISRF due to dust absorption, H$_{2}$ self-shielding etc.\ under the assumption that the gas is chemically homogeneous.
The one-zone models are run until the gas reaches chemical and thermal equilibrium. Although these calculations are extremely simplified in comparison to our full 3D models,
they have the great advantage that they are very fast, and hence it is possible to fully sample the range of number densities and column densities considered in this study with
much higher resolution than would be possible using our dynamical models. As we shall see in later sections, these simplified models prove to be helpful for interpreting the results from our full 3D runs.

\subsection{The clouds modelled in this study}
\label{sec:clouds}

\begin{figure*}
\centerline{
\includegraphics[width=3.4in]{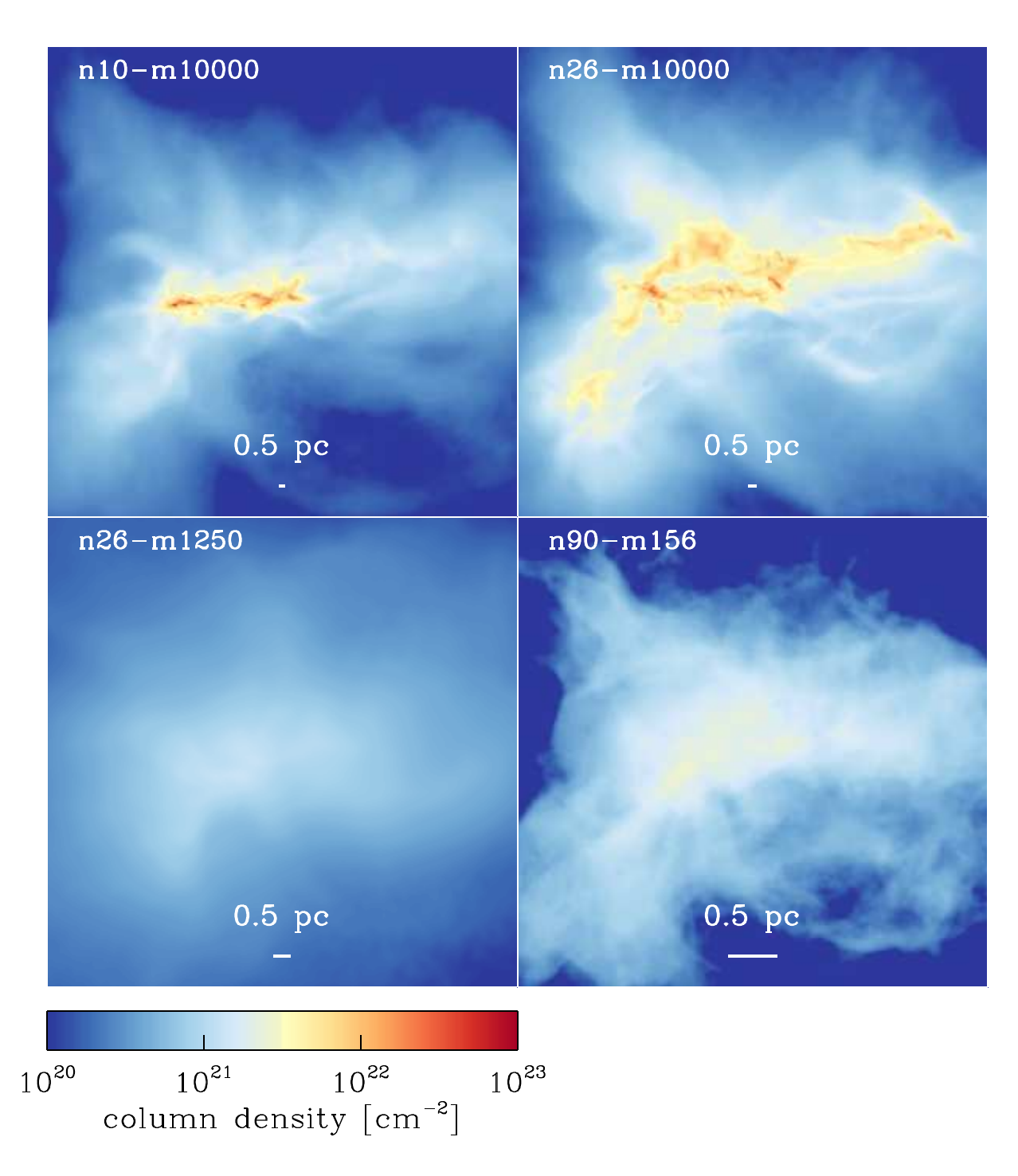}
\includegraphics[width=3.4in]{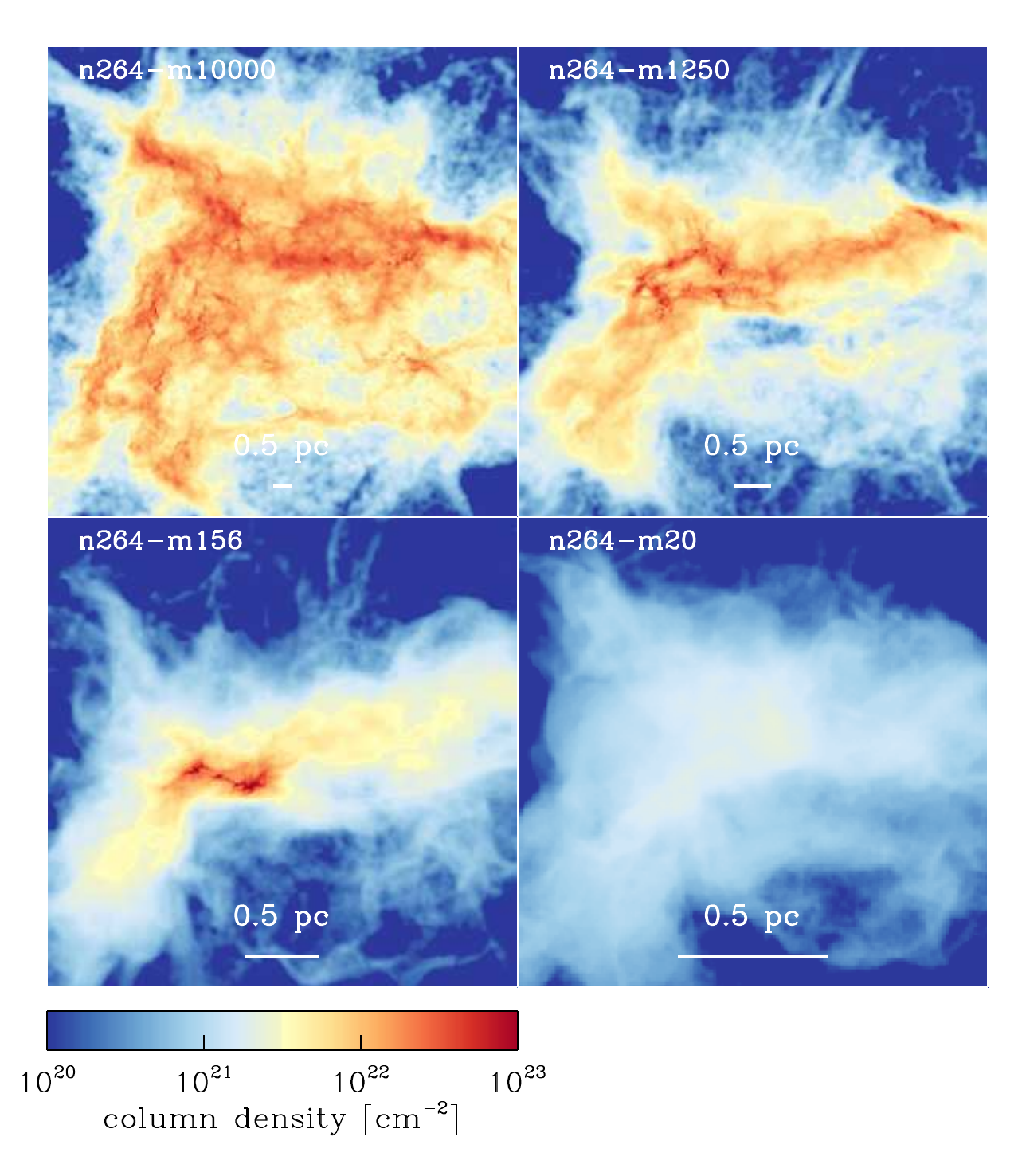}
}
\caption{Column density images from several of our low and medium density runs. Each panel has the cloud label that can be used to identify the clouds in Table \ref{sim_table}. The size of the box is chosen to match the diameter of the cloud at the beginning of the simulation. For the starless clouds, the image shows the state of the gas after a single free-fall time. The other clouds are shown just before the onset of star formation.}
\label{fig:cloudimages}
\end{figure*}

One important goal of this paper is to address a fairly simple question: if a cloud is born with a certain mass and size, will it form stars? As such, we want to cover a wide range of cloud masses and sizes to see how the conditions in the clouds change. At first glance the problem may seem trivial. At the very minimum, a cloud can only collapse if its mass exceeds the Jeans mass,
\begin{equation}
\label{mj}
m_{\rm J} = \left[ \frac{4 \pi \rho}{3} \right]^{-1/2} \left[ \frac{5 k_{\rm b} T}{2G \mu m_{\rm p}} \right]^{3/2},
\end{equation}
where $\rho$ is the mass density, $T$ is the temperature, and $\mu$ is the mean molecular weight, in units of the proton mass $m_{\rm p}$, and 
where this expression has been derived by equating the gravitational energy of a uniform density sphere to its internal thermal energy. As many of the cooling processes in molecular clouds are a function of density, one would imagine that it should be simple to determine when a cloud is bound. However the heating processes, such as H$_2$ dissociation and photoelectric emission from dust, depend on both the volume density and the column density of the cloud. This means that the temperature of the gas is not
determined purely by local conditions, but is also a function of position within the cloud. Add to that the fact that turbulence can sweep up large regions of gas to create dense (and thus cool) structures, not to mention holes through which the external radiation field can penetrate,  and we can see that the problem is potentially more complex than one would first imagine.

The suite of clouds studied in this paper are shown in the mass-radius diagram in Figure \ref{mass-radius}. For convenience, we have also plotted lines of constant column density, as well as lines of constant number density. The oft-quoted law of `constant column', or Larson's third law \citep{larson81} is also shown. Real molecular clouds do not sit exactly on this relationship, but instead display a significant scatter in their column densities \citep[see e.g.][]{heyer09,rd10}.  Since we are interested in determining the minimum column density that a cloud requires in order to form stars, we consider in this study only clouds with column densities that place them below the Larson relationship, while acknowledging that in the real ISM, there will also be a number of clouds with column densities that place them above the Larson relationship. Our ensemble of models is centred around 4 different cloud masses: 10,000 \solmasp, 1250 \solmasp, 156 \solmasp, and 20 \solmasp. 

All of the clouds are initially uniform density spheres, and as such have a distinct edge, and evolve under non-periodic gravity. At the start of the simulations, we impose a turbulent velocity field that has a power spectrum of $P(k) \propto k^{-4}$, which is left to freely decay as the simulation evolves. The initial magnitude of the turbulence is chosen such that the bulk kinetic energy is half the magnitude of the gravitational energy of the cloud. To ensure that the turbulent velocity field is modelled with the same spatial resolution in each simulation, we decided to keep the number of SPH particles fixed at 2 million for all of the clouds. We note that this means that the mass resolution (usually taken to be 100 times the mass of a single SPH particle; see e.g.\ \citealt{bb97}) varies considerably between the different simulations, from 0.001 \solmas in the 20 \solmas clouds to 0.5 \solmas in the 10,000 \solmas clouds. However, \citet{gc12a} have shown that the star formation rate in 10,000 \solmas clouds is already converged at 0.5 \solmas resolution, implying that the differing mass resolution employed here should not significantly affect the ability of the clouds to form stars. Physically, we can understand this by noting that a mass resolution of 0.5 \solmas is already sufficient to resolve essentially all of the pre-stellar cores in the cloud that go on to form stars. Improving the mass resolution beyond this point improves our ability to model fragmentation within these cores, which is important if one is interested in the initial mass function of the stars that form, but not if one merely wants to know how much mass is converted to stars per unit time. It should also be noted our choices for the sink creation density and sink accretion radius in our simulations have the effect of suppressing small-scale fragmentation within pre-stellar cores even in those simulations that would otherwise have been able to resolve this. Therefore, although the total mass of stars formed in each simulation should be a robust result, the mass function of the sink particles that form -- our best proxy for the stellar initial mass function -- will not be robust, and for that reason we do not discuss it further in this study.

\begin{figure*}
\centerline{\includegraphics[width=7.0in]{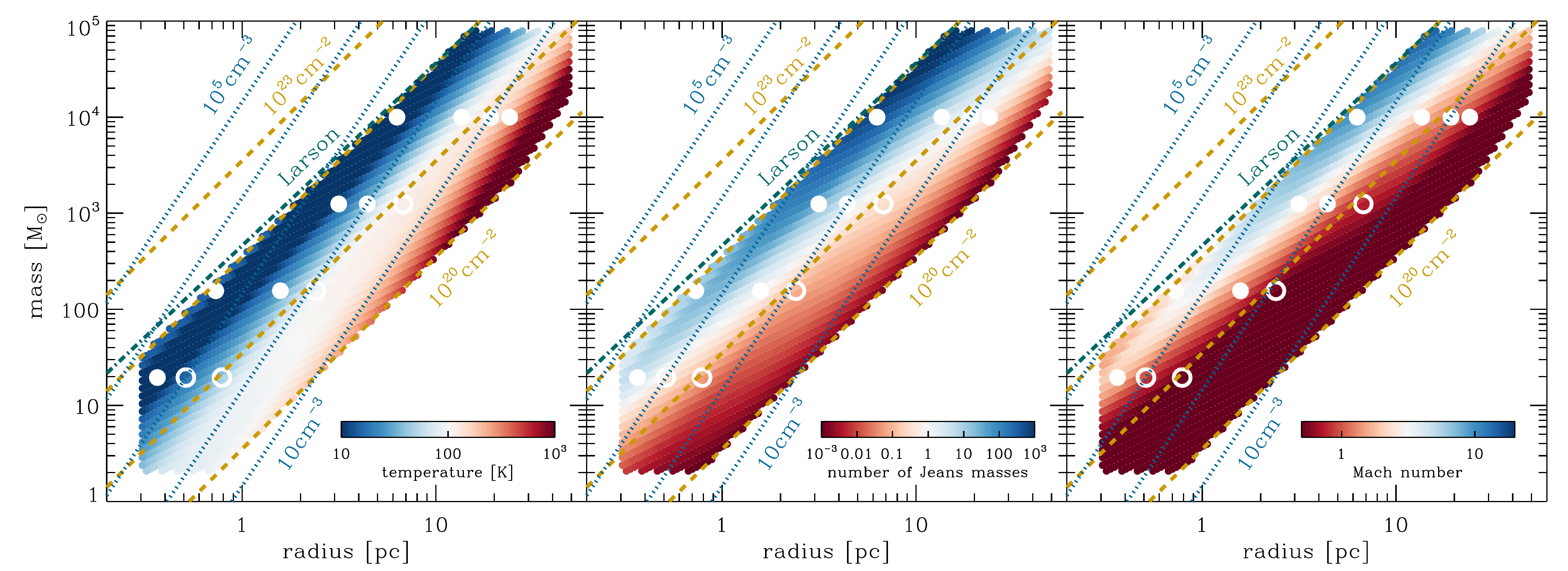}}
\caption{In the left-hand panel, the coloured region indicates the equilibrium temperatures predicted by our one-zone model, plotted as a function of the mass and radius of the clouds. The centre panel shows the number of Jeans masses, $N_{\rm J}$, contained within the clouds, while the right-hand panel shows the Mach number of the turbulence within the cloud. In both cases, these values are computed assuming that the cloud is isothermal and at the equilibrium temperature predicted by our one-zone model. In addition, when computing the Mach number, we have assumed that the kinetic energy of the turbulence is the same as in our SPH simulations, i.e.\ $E_{\rm kin} = 0.5 |E_{\rm grav}|$. In all three panels, the lines and white circles have the same meanings as in Figure~\ref{mass-radius}.}
\label{numjeans}
\end{figure*}

We fix the chemical composition of the gas by assuming that the hydrogen, helium and oxygen start in fully atomic form, and that all of the carbon is initially in the form of
C$^{+}$. We note that in reality, the conversion of hydrogen from H to H$_{2}$ will likely occur before or during the assembly of the cloud \citep{dobbs08,clark12}, and for this
reason we slightly overestimate the heating rate of the gas in the clouds, owing to the contribution made by H$_{2}$ formation heating. However, in practice this process is of
only minor importance compared to other heating mechanisms, such as the photoelectric effect, cosmic ray heating, or the influence of dynamical processes such as shocks
or adiabatic compression \citep{gc12a}. We therefore do not expect our main results to be sensitive to this choice.

Each of our simulations was run until one of two conditions was met: either 10\% of the mass of the cloud had been converted into stars, or three gravitational free-fall times (computed for the mean cloud density) had elapsed. We found that in practice, the simulations split into two categories:  either they ran for the full three fall-fall times without forming any stars, or they reached the required star formation efficiency within two free-fall times (a full discussion of the star formation rate will be given in Section \ref{sec:sfr}). Table~\ref{default} lists the time at which each simulation was halted, $t_{\rm end}$, as well as the time at which the first sink particle formed, $t_{\rm SF}$, in those clouds that were able to form stars. The end state of a sample of the simulations performed in this study can be seen in Fig. 2.

\section{Is there a column density threshold for star formation?}
\label{sec:threshold}
\subsection{Results from three-dimensional models}
The first goal of our paper is to establish how the ability of a cloud to form stars depends on its mean column density. Is there a clear column density threshold, below which 
clouds are unable to collapse and form stars, or is the situation more complicated? The results of our simulations demonstrate that although the cloud column density plays  an important role in determining whether a given cloud can form stars, the behaviour cannot be described in terms of a single column density threshold, valid for all cloud masses. This is illustrated in Figure~\ref{mass-radius}, where in addition to showing the mass and radius of our model clouds, we also indicate how successful they were at forming stars. Filled circles in Figure~\ref{mass-radius} correspond to star-forming clouds, while open circles denote ``sterile'' clouds that do not form any stars within three gravitational free-fall times. We see that if we increase the cloud radius while keeping the cloud mass fixed, which corresponds to reducing both the volume density and the column density of the cloud, then eventually our clouds become sterile. For our cloud models with masses of 156 \solmasp~and 1250 \solmasp, this transition occurs at a column density of around $10^{21}\,$cm$^{-2}$, and it is tempting to identify this as a threshold column density. However, our lowest mass clouds become sterile at higher column densities, while our highest mass clouds are still able to form stars at somewhat lower column densities, suggesting that there is no single threshold value that is valid for all cloud masses. 

\subsection{Results from one-zone models}
To get a better understanding of the physics involved, we performed a large number of one-zone models of `clouds' (defined by a single mass, radius, volume density, and column density) to compute an estimate of the temperature in the regime of interest in the mass-radius diagram. These temperatures can then be used in conjunction with the cloud's volume density to compute a mean Jeans mass. From this, we can estimate the number of Jeans masses in the cloud (i.e.\ $N_{\rm J} = M_{\rm cloud}/m_{\rm J}$), which should give a good indication of the likelihood that a given cloud will be capable of star formation. The results from such an analysis are shown in Figure \ref{numjeans}. Also shown is the mean Mach number of the turbulence that one would expect to find in the clouds, computed using the assumption that the turbulent kinetic energy is equal to half of the gravitational energy, as in the initial conditions for our 3D models.

From the left-hand panel of Figure~\ref{numjeans}, which shows the temperatures derived from our one-zone model, we see that for clouds with column densities greater than $N \sim 10^{21}\,$cm$^{-2}$, lines of constant temperature roughly follow lines of constant column density. This is due primarily to the influence of photoelectric heating, which is the dominant heat source in low $N$ clouds, but which becomes much less effective once the cloud becomes optically thick to photons with $h\nu > 6 \: {\rm eV}$.  
In the unshielded regime, the equilibrium temperature of the cloud is around 100~K or more, but once the cloud becomes able to shield itself against the incident photons and reduce the efficacy of the photoelectric heating, its temperature drops to around 20~K, driven by C$^{+}$ cooling. Additional cooling occurs once the visual extinction exceeds a few magnitudes and the cloud becomes capable of maintaining a large CO fraction. At even higher column densities, the CO lines become optically thick and the equilibrium temperature rises slightly, but it remains relatively small, of order 10~K.

\begin{figure*}
\centerline{\includegraphics[width=7.0in]{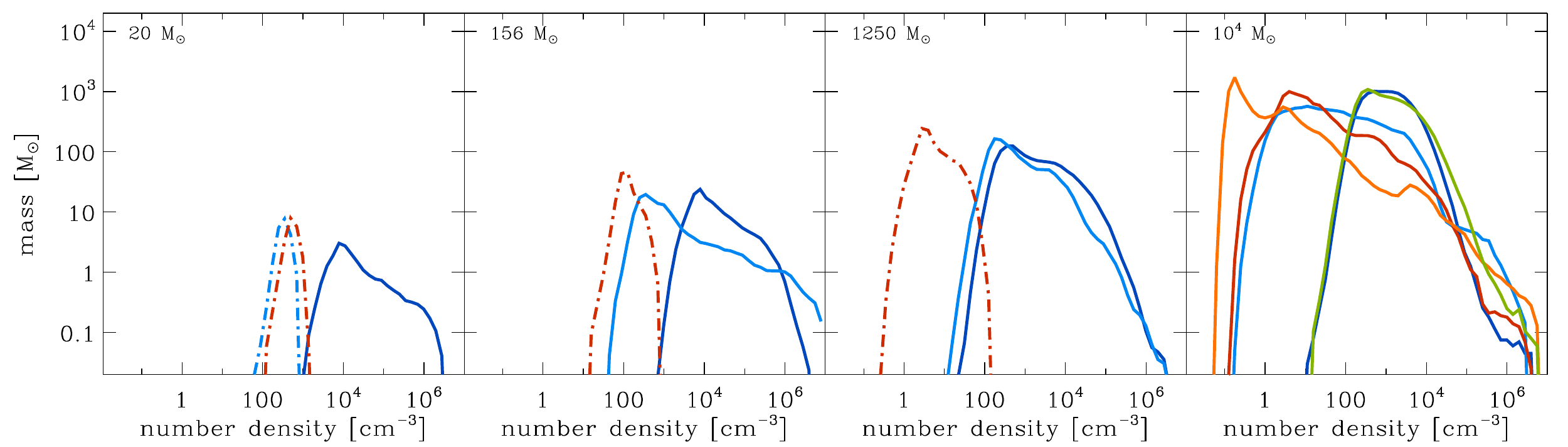}}
\centerline{\includegraphics[width=7.0in]{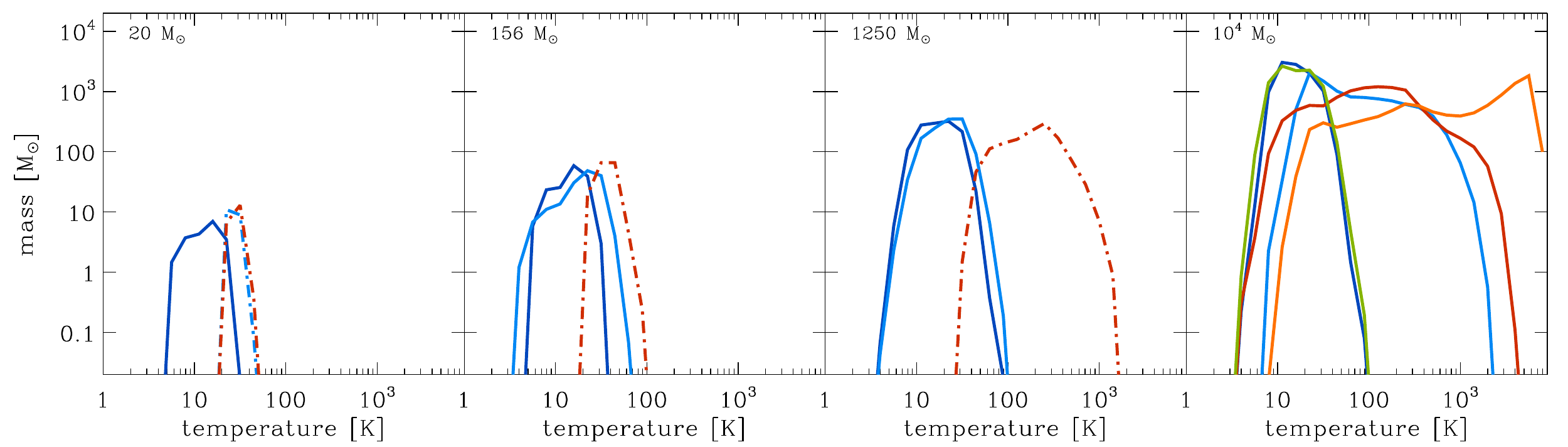}}
\centerline{\includegraphics[width=7.0in]{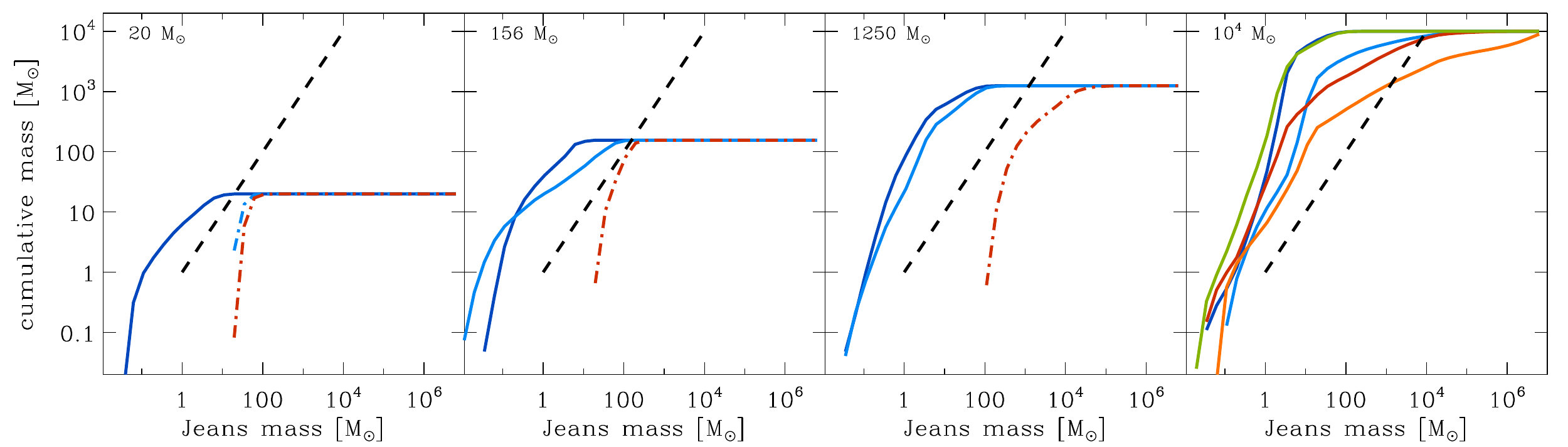}}
\caption{The top and middle rows of this figure show, respectively, the mass-weighted PDFs of number density and temperature for the clouds in our study. 
Star-forming clouds are shown as solid lines, while those that are sterile (i.e.\ not star-forming)
are depicted with dot-dashed lines. Each individual panel shows clouds of the same mass. To distinguish between clouds, we have taken the colours of the lines to match those of the symbol colours given in Table~\ref{sim_table}. Starless clouds are depicted after one initial free-fall time, while star-forming clouds are shown immediately before the onset of star formation. The bottom row shows the cumulative mass distribution of the Jeans masses. Assuming the cumulative mass is spatially coherent, curves sitting above the black-dashed line show clouds (or parts of the cloud) that should be able to collapse -- at least according to the classical Jeans analysis. As such, one would expect star-forming clouds to sit (at least in part) above the black-dashed line, while the sterile clouds should sit completely below the line. The figure demonstrates that this is indeed the case.}
\label{fig:pdfs}
\end{figure*}

\begin{figure*}
\centerline{
	\includegraphics[width=3.4in]{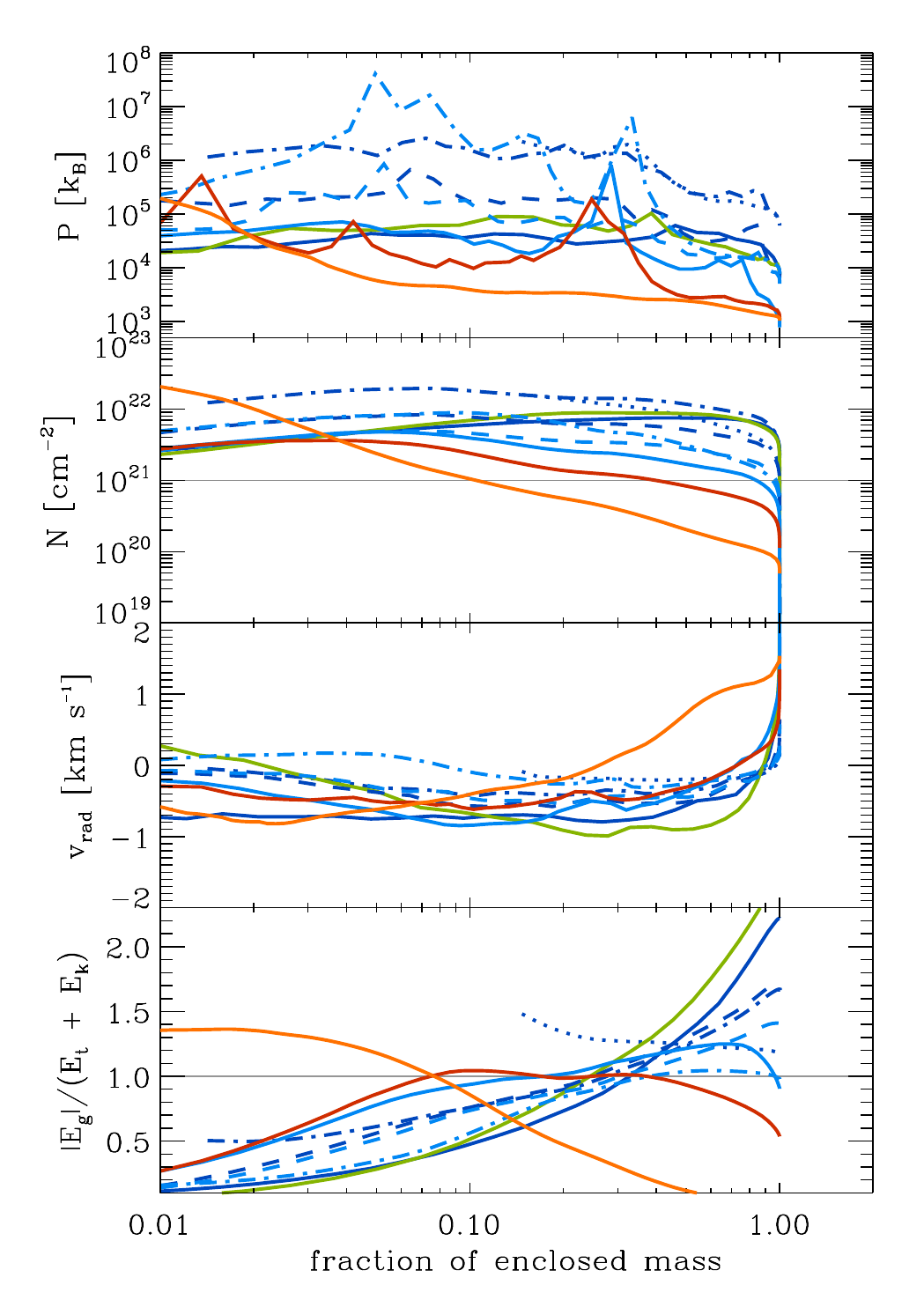}
	\includegraphics[width=3.4in]{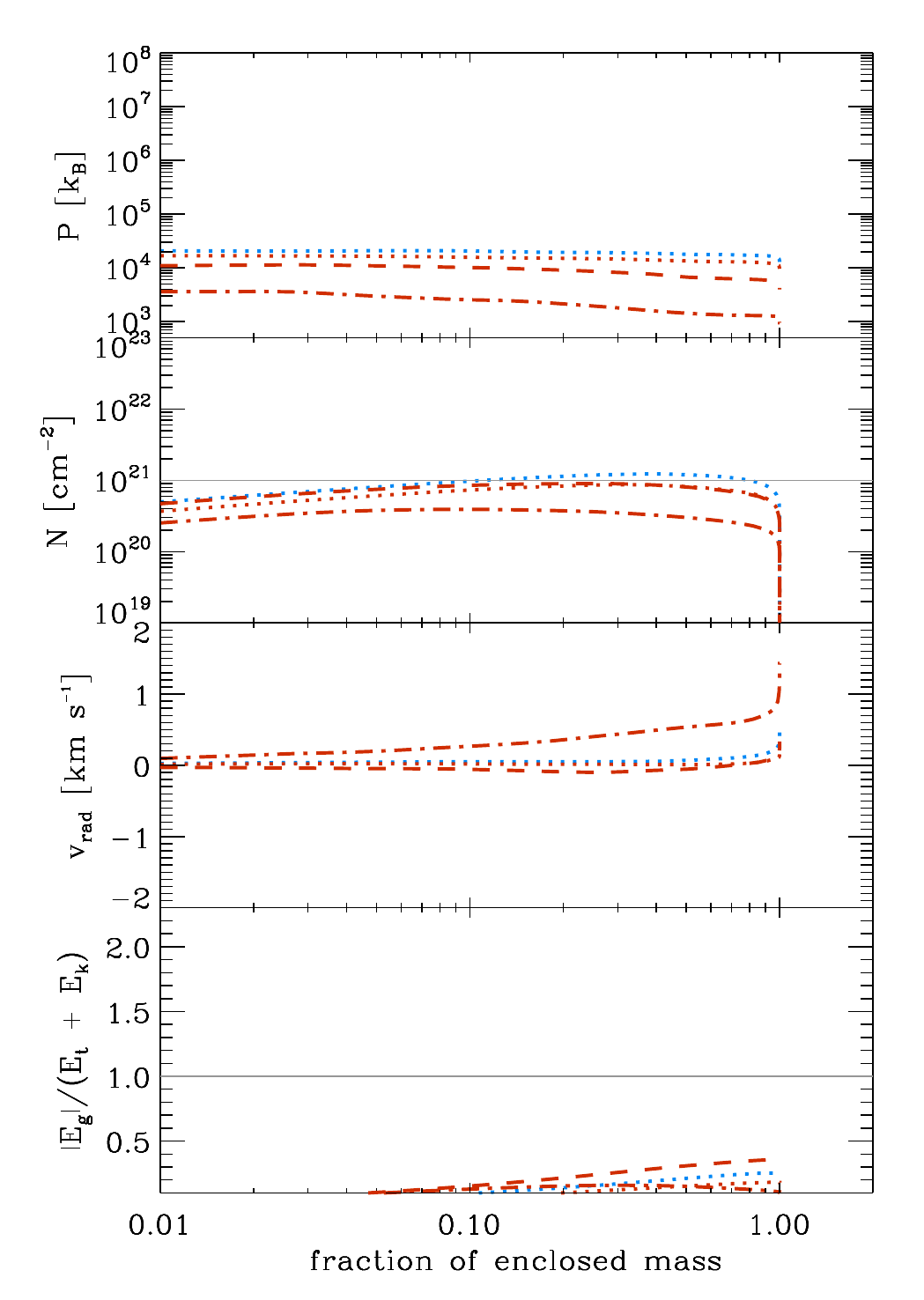}
}
\caption{ We show here several properties of the clouds as a function of the radially enclosed mass.  The left-hand panel shows the star-forming clouds at a time $t_{\rm SF}$, while the right-hand panel shows the sterile clouds at time $t_{\rm ff}$. In the majority of the cases, the cloud centre is defined to be the centre of mass (and velocity). However, in the case of the very low density clouds (runs {\it n5-m10000} and {\it n10-m10000}), the centres were taken to be the densest SPH particle, since the star-forming regions were located some distance away from the centre-of-mass of the clouds. The pressure is calculated from the mass-weighted means of the number density and temperature in each radial bin. The column density shown here is computed via $N = M_{\rm enc} / \pi R^2$, where $M_{\rm enc}$ is the enclosed mass at radius $R$.  The radial velocity, $v_{\rm rad}$ denotes the mass-weighted mean radial velocity in the shell. Finally, the gravitational energy as a function of radius is given by $E_{\rm g}(R) = G\, M_{\rm enc}\, m(R) / R$, where m(R) is the mass of the shell at radius $R$, and the kinetic ($E_{\rm k}$) and thermal ($E_{\rm t}$) energies are computed from the running sum of the particle energies in each shell (with the velocities taken relative to the cloud's centre-of-velocity). Note that due to the non-spherical nature of the clouds, these radial properties are approximate. However, we find that they are accurate  to within a factor of a few and give a fair representation of the overall trends within the clouds. All colours and line styles are the same as those listed in Table \ref{sim_table}.}
\label{fig:cloudprofiles}
\end{figure*}

If we now look at the number of Jeans masses that we expect to find in the clouds, illustrated in the centre panel of  Figure~\ref{numjeans}, we see that lines of constant
column density do not correspond to lines of constant $N_{\rm J}$. Instead, $N_{\rm J}$ is significantly smaller for low mass, high volume density clouds, than for higher
mass, lower volume density clouds with the same column density. This behaviour is actually fairly simple to understand. An idealised spherical cloud with volume density 
$n$ and radius $R$ has a mass that scales as $M \sim n R^{3} \sim N^{3} n^{-2}$, where $N \sim n R$ is the column density of the cloud, measured between the edge and
the centre. If the temperature of the gas is $T$, then the Jeans mass within the cloud scales as $m_{\rm J} \sim T^{3/2} n^{-1/2}$. Therefore $N_{\rm J} = M / m_{\rm J}
\sim N^{3} T^{-3/2} n^{-3/2}$. If we assume that $T$ remains almost constant as we vary $n$ provided that we keep the column density $N$ fixed,  then we find that at
constant column density, $N_{\rm J} \sim n^{-3/2}$. In other words, our cloud contains fewer Jeans masses when its density is high than when its density is low. 
This behaviour occurs because although $m_{\rm J}$ decreases with increasing $n$, the cloud mass decreases much more rapidly: for a cloud at fixed column density,
$M \sim n^{-2}$.

An obvious requirement for star formation to occur is that at least part of our cloud must be unstable to gravitational collapse. Supersonic turbulence can create 
unstable regions within clouds that are themselves gravitationally unbound \citep[see e.g.][]{clark05,dobbs11}, but nevertheless, we expect clouds with $N_{\rm J} \gg 1$ 
to be much more effective at forming stars than clouds with $N_{\rm J} \ll 1$. If we therefore adopt the requirement that $N_{\rm J} > 1$ in order for efficient star 
formation to occur, then
our analysis above demonstrates that this condition will be satisfied at different column densities for different cloud masses. In other words, there is no single
column density threshold -- the threshold value is a function of the cloud mass, in agreement with the results plotted in Figure~\ref{numjeans}. However, we also
see that $N_{\rm J}$ depends strongly on the value of the column density, particularly once the cloud starts to shield itself from the influence of the interstellar 
radiation field. We would therefore not expect the threshold value to vary by much as we vary the cloud mass. Again, this is in good agreement with the results
shown in Figure~\ref{numjeans}, which show that the threshold column density for efficient star formation varies by less than an order of magnitude as we vary 
the cloud mass by almost three orders of magnitude. 

In general, we find that the simple Jeans analysis from the one-zone models does a good job of predicting regions of the mass-radius diagram in which star formation should 
be possible. However, if we look at the most massive clouds in our study, we see that the predictive power of the one-zone model begins to break down. All of our simulated
$10^{4} \: {\rm M_{\odot}}$ clouds form stars, even though the one-zone model predicts that the lowest density example (run {\em n5-m10000}) 
should have $N_{\rm J} < 1$. One important
reason that the one-zone model breaks down in this case is that our assumption that the cloud has an approximately isothermal temperature distribution becomes increasingly
inaccurate as we move to lower mean densities. In our dense, low column density clouds, the difference between the temperature at the edge of the cloud and in the
mildly shielded region at the cloud centre is fairly small, as the plots in Figure~\ref{fig:pdfs} demonstrate. As we decrease the cloud density, however, the 
temperature difference grows significantly. This occurs because at densities below $n \sim 10 \: {\rm cm^{-3}}$, the equilibrium temperature becomes a very steep function of density
\citep[see e.g.][]{wolf95}. Small differences in the thermal pressure (due, for instance, to the small amount of shielding that is present at the cloud centre) lead to density 
gradients that then create much larger differences in the temperature and pressure. We shall look at this in more detail in Section \ref{sec:conditions}.

\subsection{The role of turbulence}
Another weakness in our one-zone models is that they 
do not allow us to account for the effect of turbulence within the clouds, which can create bound sub-structures even in clouds with
$N_{\rm J} < 1$ by triggering thermal instability \citep{Heitsch2006, vs07, banerjee09}. We can get some idea of the regions in the mass-radius diagram in which we might expect turbulence to be important by computing
the expected turbulent Mach number for each cloud. The resulting values are plotted in the right-hand panel of Figure~\ref{numjeans}. We see that in practice, clouds with 
$N_{\rm J} \sim 1$ that are initially close to virial equilibrium, as in this case, will have a turbulent Mach number of around 1. This is to be expected: clouds with only a single 
Jeans mass have thermal and gravitational energies that are equal, and clouds that are close to virial equilibrium have kinetic energies that are comparable to their gravitational 
energy, so it follows that the turbulent kinetic energy and thermal energy of these clouds will also be comparable. 
 
The turbulence within the clouds contributes to their global stability, but on smaller scales leads to enhanced fragmentation \citep{mlk2004}. It plays an important role in driving star formation in our large, low-density clouds, where the amount of mass entrained by the turbulent flows is large.  In these low density clouds, which have mean densities that place them in the unstable regime between the warm neutral medium and the cold neutral medium, the relatively small density perturbations associated with the transonic turbulence are nevertheless sufficient to cause large changes in the temperature. Regions that cool substantially are further compressed by their surroundings, with the result that the local Jeans mass decreases within these regions by a very large amount. In the high density gas clouds, this process is not possible: the gas is already cold and the turbulent flows entrain less mass. Additionally, it should be noted that the turbulent velocity fields that are used to impose the initial turbulence in the clouds in our simulations are created with a ``natural'' mix of 2:1 solenoidal to compressive modes. If one were to adopt a higher compressive contribution (due, for example, to orbit crowding within spiral arms, or to compression by a supernova remnant or H$\,${\sc ii} regions), then more high density gas would be created, potentially enhancing the prospects for star formation \citep{fks08,fk12}.

One possible concern regarding our treatment of the turbulence within the clouds is that we do not drive the turbulence with a continuous input of kinetic energy, but instead simply allow the initial turbulent kinetic energy to decay. The question of whether this is a good description of the behaviour of real Galactic molecular clouds is somewhat contentious. Supersonic turbulence is known to decay with a characteristic timescale comparable to the turbulent crossing time, $t_{\rm cross} \sim R / v_{\rm turb}$ \citep{maclow99}. If typical molecular cloud lifetimes are much longer than $t_{\rm cross}$, then the fact that we observe supersonic, turbulent linewidths in all molecular clouds implies that some process must be replenishing the turbulent kinetic energy. On the other hand, if clouds only live for 1--2 crossing times \citep[see e.g.][]{elmegreen00}, then there is no need to assume that the turbulence is being continually driven.

In our star-forming clouds, the onset of star formation typically occurs after approximately a single turbulent crossing time, at which point a significant fraction of the initial turbulent kinetic energy remains within the cloud. It therefore is highly likely that these clouds would still form stars even if we were to drive the turbulence within them, particularly since previous studies of the effects of driven turbulence on the star formation rate show that it completely suppresses star formation only when driven on very small scales \citep[see e.g.][]{khm00}.

In the case of our sterile clouds, it is possible that part of their gas might be induced to collapse if we were to continue to drive the turbulence within them. However, we consider this to be unlikely: the cooling time of the gas in these clouds is much shorter than the turbulent crossing time, and so if the turbulence was going to be sufficient to trigger extensive cooling and collapse, we would expect this to occur within the first 1--2 dynamical times, and hence we should see signs of this occurring within our own simulations.

To sum up, our simulations suggest that the cloud mass needs to be $\ga 10^4\,$\solmasp~in order for stars to form when the total cloud-averaged
column density is below $10^{21} \,$cm$^{-2}$. In lower mass clouds, the required column density is a factor of a few larger. Nevertheless, even in this case, the required column densities are significantly smaller than the threshold value of around $10^{22} \,$cm$^{-2}$
proposed by \citet{heid10}. This suggests that the observed ``threshold'' in local star-forming regions is probably a tracer of the star formation activity, rather than a physical condition that must be met before star formation can occur, as we discuss further in Section~\ref{sec:ak08} below.

\begin{figure}
\centerline{
\includegraphics[width=3.3in]{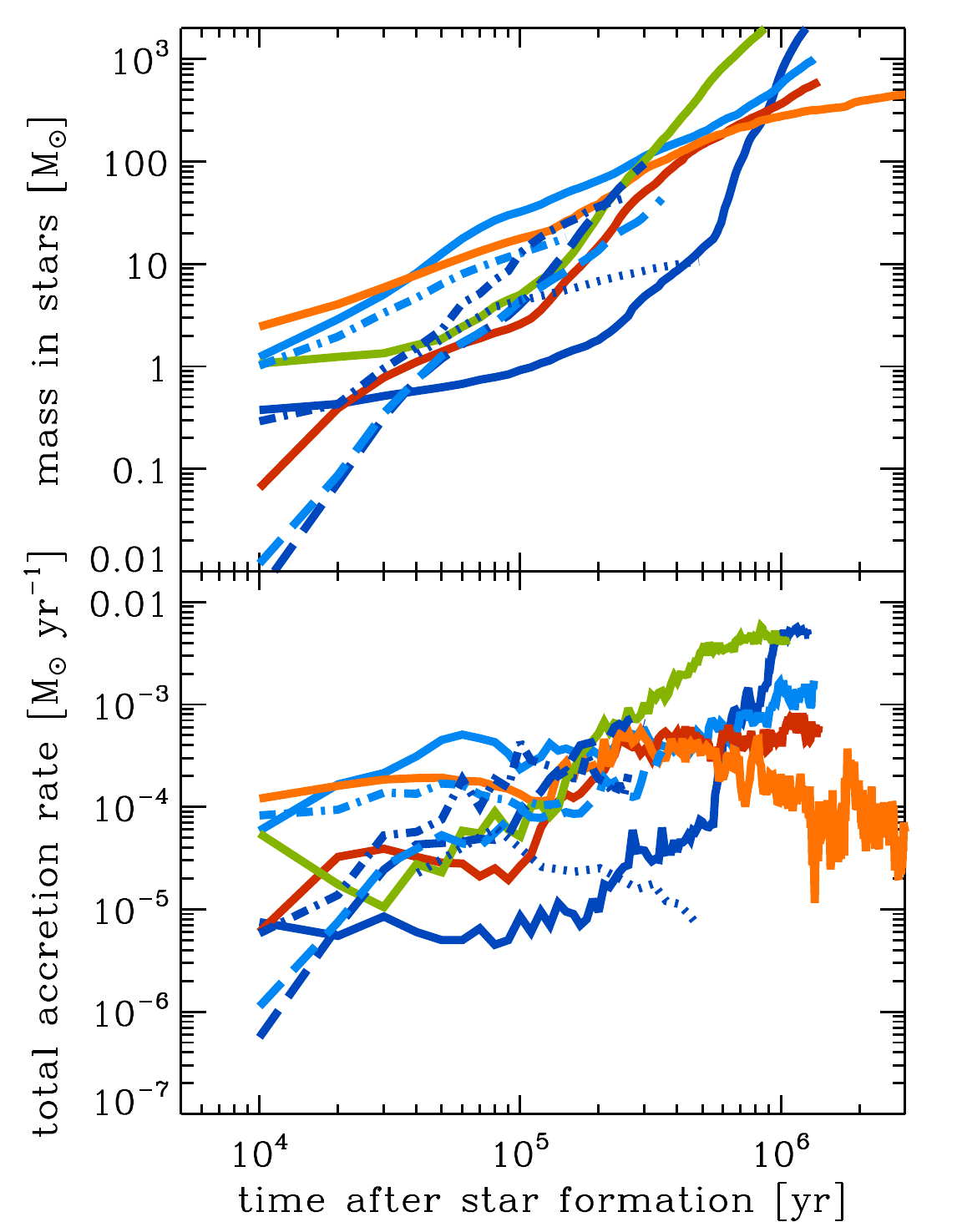}
}
\caption{Mass in sink particles as a function of time, plotted for each of the star-forming clouds. We assume that all of the mass accreted by the sinks is converted into stars,
and hence this plot shows how the mass in stars increases with time in the clouds. The different line styles are used to denote the mass of the clouds, with the following convention: solid lines correspond to clouds with 10,000 \solmasp; dashed lines correspond to clouds with 1250 \solmasp; dot-dashed lines correspond to clouds with 156 \solmasp; and dotted lines correspond to clouds with 20 \solmasp. The colours of the lines within each mass-group are consistent with those given in Figure~\ref{mass-radius}.}
\label{fig:mstar_time}
\end{figure}

%
%
%
%
%
%
%
\section{Physical conditions in the clouds}
\label{sec:conditions}

Although the one-zone calculations seem to be able to predict where star formation can occur on the mass-radius diagram, they say nothing about the internal structure of the clouds, and the range of densities and temperatures that one should expect. To this end, in Figure~\ref{fig:pdfs} we look at the temperature and density distributions in our 3D model clouds. As in all of our figures, star-forming clouds are depicted just before the onset of star formation (within $10^4$ yr of the first sink particle forming -- set by the rate at which we write snapshots), while the sterile clouds are depicted after one initial free-fall time. However, we note that in the star-forming clouds, the onset of star formation tends to occur at around a free-fall time (see Table~\ref{sim_table}), and so all the clouds are shown at a roughly equivalent evolutionary point.

The probability density functions (PDF) describing the number density in each of the clouds are shown in the top row of panels in Figure~\ref{fig:pdfs}. We see that clouds that form stars (the solid lines) have significantly broader PDFs than their sterile counterparts (the dot-dashed lines). This is a direct consequence of the gravitational collapse that is occurring on small scales within the star-forming clouds. This leads to the density PDFs of these clouds developing 
power-law-like features at the high density end, similar to those reported by a number of previous studies \citep[see e.g.][]{klessen00,slyz05,fed08,vs08,collins11,krit11,fk13}.

If we now look at the temperature PDFs, shown in the middle row of panels in Figure~\ref{fig:pdfs}, we see that it is more difficult to distinguish between star-forming and sterile clouds. 
The width of the temperature PDFs is determined primarily by the mean density of the cloud, rather than by whether or not the cloud is collapsing. Our massive clouds,
which have relatively low mean densities, have broad temperature PDFs, because at low densities, the equilibrium gas temperature is a strongly varying function of 
density. On the other hand, our lower mass, denser clouds have narrower temperature PDFs, because at higher densities, the dependence of temperature on density
is less pronounced. Nevertheless, even in this case it is clear that our clouds are not isothermal, but rather exhibit a range of different temperatures. 

We also see that the mean temperature of the clouds is not a good diagnostic of whether or not they will form stars. The sterile clouds in runs {\it n264-m20} and {\it n1000-m20} have temperature distributions that peak around 30~K, but so does the star-forming cloud in run {\it n264-m156}, while the star-forming clouds in runs {\it n5-m10000} and {\it n10-m10000} have distributions that peak at much higher temperatures. The best indication of star formation comes from the low temperature end of the PDF. All of our star-forming clouds contain gas with $T < 20$~K, often in large quantities. On the other hand, none of our sterile clouds contain gas that is this cold. We can explain this difference rather easily. In order to produce gas with  $T < 20$~K, we require two things: a relatively high extinction, so that the gas is well-shielded from photoelectric heating, and a high CO fraction, as C$^{+}$ cooling alone has difficulty in cooling the gas below 20~K \citep{gc12a}. Both conditions are satisfied within the dense, gravitationally collapsing structures that form within our star-forming clouds, but are generally not satisfied within the lower density, gravitationally stable gas that we find within our sterile clouds. 

The bottom row of panels in Figure~\ref{fig:pdfs} shows the amount of bound gas that is present in each cloud. The panels show the cumulative amount of mass in the cloud that has a Jeans mass below the value plotted on the x-axis. The black dashed line shows the linear relationship that denotes how much of the cloud is gravitationally bound (at least in terms of the thermal support), such that the amount of mass lying above this line represents the amount of gas that is potentially available for star formation. We see that all of the sterile clouds lie below the dashed line, implying that they have no regions within them that are gravitationally bound. This result is consistent with the outcome of the Jeans analysis that we performed using the results from the one-zone cloud models, and provides a natural explanation for why these clouds fail to form stars. On the other hand, we see that in most of the star-forming clouds, essentially all of the gas is gravitationally unstable and hence is potentially able to form stars. The one exception is the very diffuse {\em n5-m10000} cloud, in which most of the gas is gravitationally stable, with only around 10\% being unstable and available for star formation.

In the massive star-forming clouds, which have broad distributions of both temperature and density, it is also interesting to examine the properties of the gravitationally bound parts of the cloud. Figure~\ref{fig:cloudprofiles} shows how several properties of the clouds vary as a function of the (radially) enclosed mass.  We see that the regions of the cloud that are gravitationally bound, and thus able to collapse to form stars, have a radially averaged column density of around $10^{21} \rm cm^{-2}$.  So, while the cloud as a whole can have a mean column density below this value and yet still be capable of forming stars, the regions in which the stars will actually form always lie above this column density.

In light of Figure~\ref{fig:cloudprofiles}, we should be careful with our definition of ``cloud''. In the case of the higher volume and column density simulations, our ``clouds'' are roughly analogous to the real clouds that we observe; local clouds are also defined in the literature based on column densities derived from extinction mapping, not only from their CO emission, and can easily reach values as low as $10^{21} \rm cm^{-2}$ at their boundaries. However, our simulations with very low column and volume densities are perhaps better interpreted as probing a typical patch of the Milky Way ISM: in general, the ISM is gravitationally stable, but it possible to have regions that can cool sufficiently to form stars. What these results show is that $10^{21} \rm cm^{-2}$ represents an effective threshold below which the ISM is in general gravitationally stable -- at least for solar neighbour values of the ISRF and CRIR. For convenience, we will retain the term ``cloud'' when referring to the simulations, but we caution the reader to consider that this term is not being used in the standard observational definition (i.e.\ our clouds can contain a mixture of warm and cold H{\sc i} gas {\it and} warm and cold H$_2$). 

\section{Characterising the star formation in the simulations}
\label{sec:sfr}

\begin{figure}
\centerline{
\includegraphics[width=3.3in]{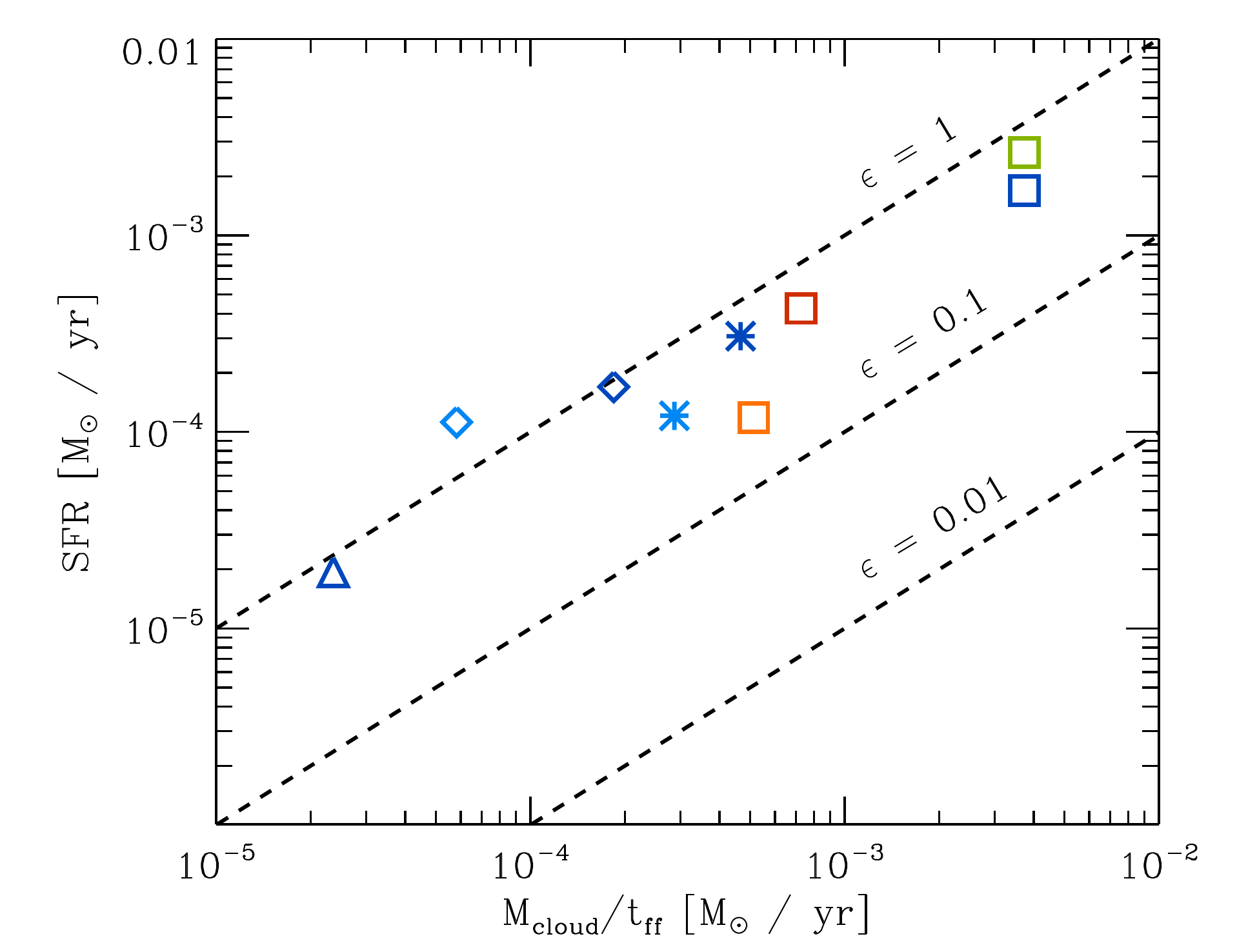}
}
\caption{The SFR in our simulated clouds as a function of $M_{\rm cloud} / t_{\rm ff}$. Lines depict constant $\epsilon$ as defined in Equation \ref{eq:sfetff}. Symbol shapes and colours are the same as those given in Table \ref{sim_table}.}
\label{fig:sfr_tff}
\end{figure}

We have seen that clouds with a wide range of different masses and densities are capable of forming stars. However, an important question remains: once these clouds start to form stars, how quickly does the conversion of gas into stars proceed? In what follows, we first discuss the different ways to measure the star formation rate in the clouds, before going on to discuss specific comparisons to our models. 

\subsection{The star formation rate}
There are two relatively straightforward ways to measure the star formation rate (SFR) in our simulations. The first is to ask how long the cloud takes to convert a given mass into stars, with the clock starting at the point at which the cloud is ``assembled''. In our case, as our clouds are pre-assembled at the start of the simulation, this amounts to measuring the time from the beginning of the simulation. This star formation rate is then defined by $M_{\rm stars}/t_{\rm end} = f_{\rm SFE}M_{\rm cloud}/t_{\rm end}$, where $M_{\rm stars}$ is the mass in stars, $t_{\rm end}$ is the time taken to form these stars, and $f_{\rm SFE}$ is the star formation efficiency, defined here as $f_{\rm SFE} \equiv M_{\rm stars} / M_{\rm cloud}$, where $M_{\rm cloud}$ is the initial mass of the cloud. 

The second method, and the one that we will adopt in much of the analysis presented in this paper is simply to average the accretion rate of the sink particles over a given period in the simulations after the formation of the first sink particle. Although the first method is  perhaps a more accurate description of how the star formation proceeds in these simulations, it is difficult to rectify with the methods used to observationally estimate the SFR in real clouds, which attempt to measure the SFR over the period in which the clouds are actively forming stars. As such, the second method is more appropriate. We will refer to this as simply the SFR, or the ``measured sink SFR'' in the rest of this paper. 

In the top panel in Figure \ref{fig:mstar_time} we plot the mass in sink particles as a function of time. The period shown covers the point at which the first sink particle forms to the point at which the clouds have converted roughly 10 percent of their mass into sink particles. The clouds are depicted in Figure \ref{fig:mstar_time} using our standard line styles and colours. The bottom panel shows the total accretion rates in the clouds as function of time.

We see that the growth of the sink particle population covers a wide range of behaviours. In some cases, such as the two {\em n264-m10000} clouds (i.e. the two different turbulent seeds), the mass accretion rate accelerates with time, while in others, such {\em n5-m10000} and {\em n2640-m20} the accretion rates fall towards late times. Others, for example {\em n10-m10000}, seem to plateau. However, what is important for this section is that by the point at which we terminate the simulations, the type of accretion behaviour -- accelerating, decelerating, plateauing -- is well characterised. Note also that in the case of the higher mass clouds, the stellar populations would at this point contain a significant fraction of high-mass stars, whose feedback processed are not captured in this study.

Given these considerations, we calculate the SFR in this study based on the mean accretion rate over the period from the onset of star formation, to the point when 10 percent of the mass of the cloud is in sink particles. We note that the values obtained by averaging over the total sink particle accretion rates in this manner are within a factor of two of the value of the SFR one would obtain simply taking $M_{\rm stars}/(t_{\rm end} - t_{\rm SF})$, where $t_{\rm SF}$ is the time at which the first sink particle forms. We have also verified that the resulting SFRs are not particularly sensitive to our decision to compute them at the point when 10\% of the cloud mass has been converted to stars. SFRs for the clouds computed at the point when only 5\% of the cloud mass has been converted to stars generally agree with our quoted values to within 10\%, with larger differences of a factor of two seen only in the runs in which star formation is strongly accelerating. Nevertheless, it should be clear to the reader from Figure \ref{fig:mstar_time} that the trends that we will discuss below in Section \ref{sec:sfrprops} will not hold at very early times in the cloud evolution, when little of the mass has been converted to stars. It is worthwhile keeping this point in mind when the considering our results. 

\begin{figure*}
\centerline
  {
    \includegraphics[width=3.3in]{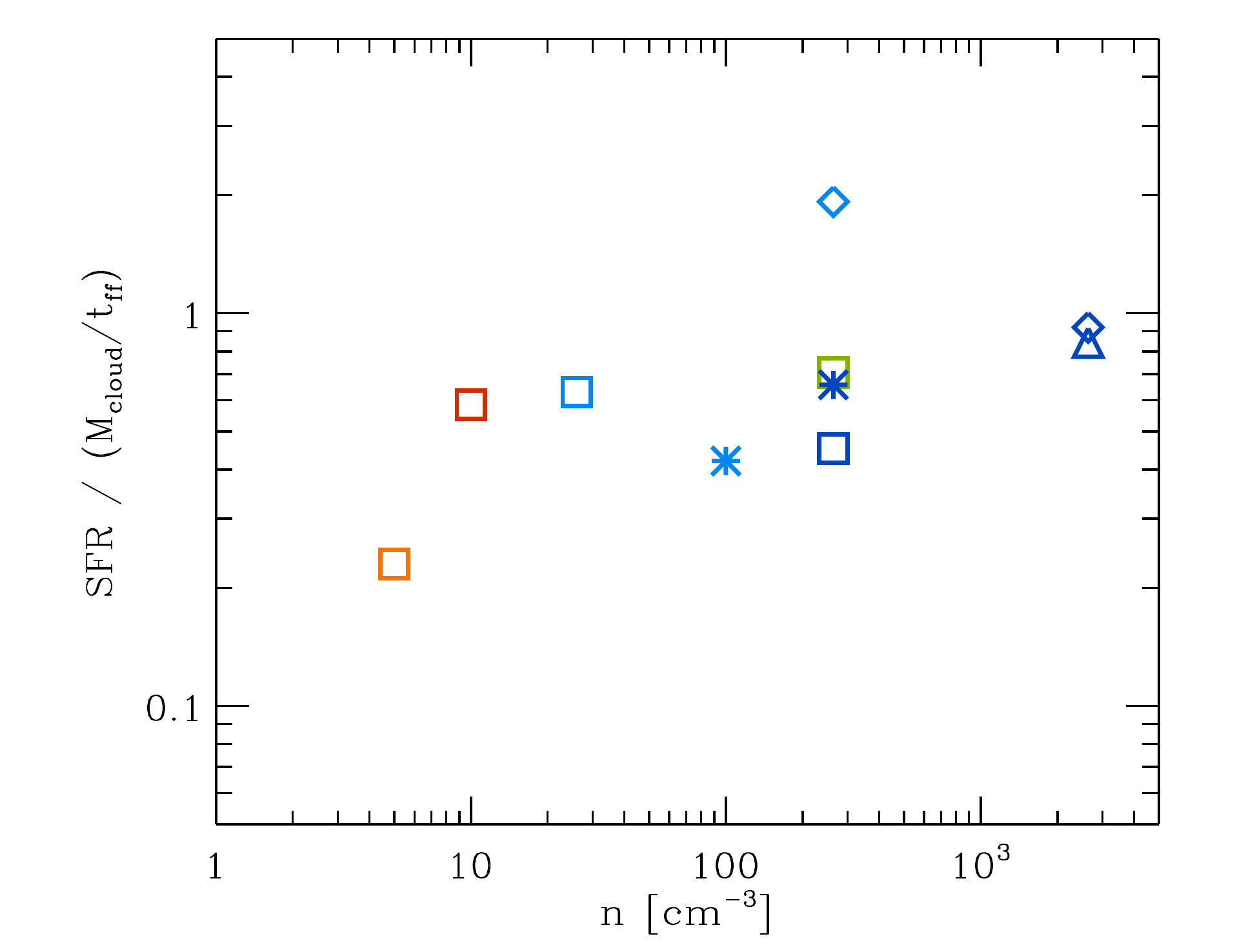}
	\includegraphics[width=3.3in]{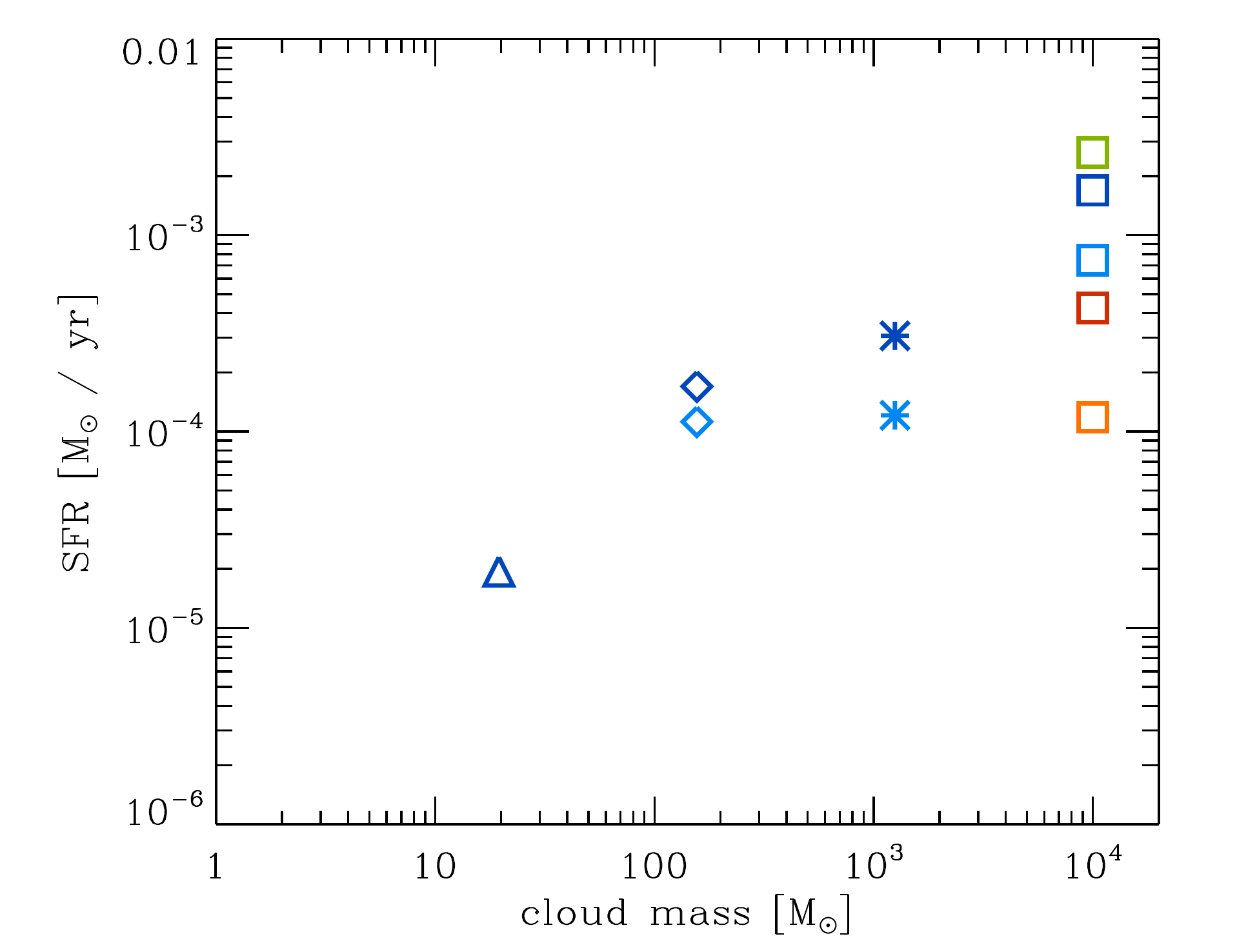}
  }
\centerline
  {
    \includegraphics[width=3.3in]{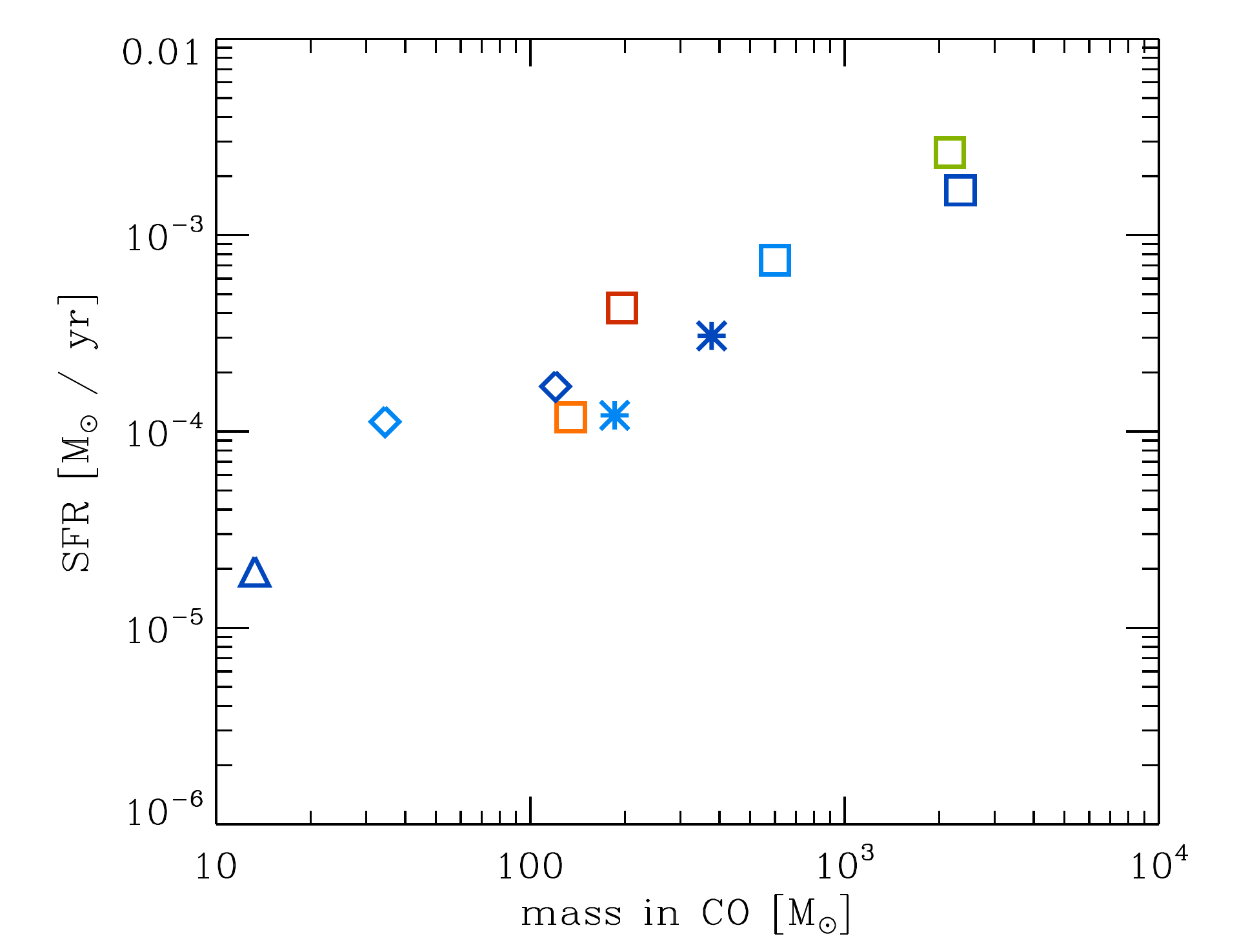}
	\includegraphics[width=3.3in]{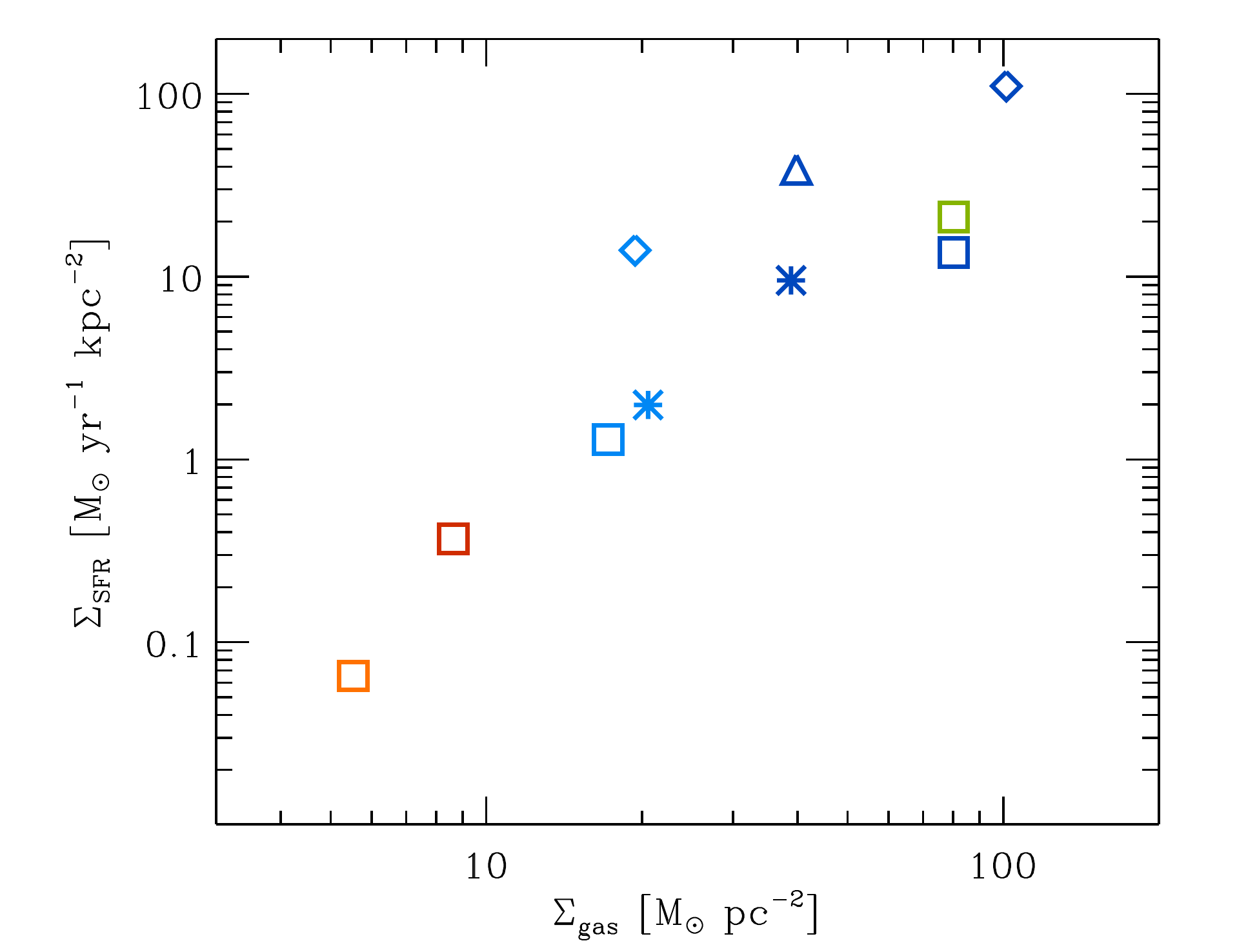}
  }
\caption{\label{sfrclouds} The panels show how the mean star formation rate (SFR) varies with cloud parameters. See Section \ref{sec:sfrprops} for more details. Symbol shapes and colours are the same as those given in Table \ref{sim_table}}
\end{figure*}

\subsection{The dependency of the star formation rate on the cloud properties}
\label{sec:sfrprops}

It has long been suggested that the star formation rate in a giant molecular cloud (GMC) can be described as some fraction $\epsilon$ of the cloud's mass $M_{\rm cloud}$ collapsing into stars on the cloud's mean free-fall time $t_{\rm ff}$,
\begin{equation}
\label{eq:sfetff}
{\rm SFR} = \frac{\epsilon \, M_{\rm cloud}}{t_{\rm ff}}.
\end{equation}
Originally, the idea became popular as an explanation of the Kennicutt-Schmidt law \citep{schmidt59,k98}. However, it has seen a recent revival as part of models of 
turbulently-regulated star formation (e.g.\ \citealt{km05,kt07,PN11, HC2011}),
which have attempted to provide a theoretical framework for predicting the value of $\epsilon$. With these later models, it has become common to express the star formation rate in terms of $\epsilon$, the fraction of the cloud converted to stars per free-fall time.\footnote{This quantity is often referred to as the star formation efficiency per free-fall time, SFE$_{\rm ff}$ \citep{km05}, although note that their definition is slightly different from that given here \citep{fk12}.}
\citet{fk12} showed that for the definition of $\epsilon$ given above, $\epsilon$ should vary substantially with the properties of the cloud, but in general should be in the range of 0.1 to 1 (although in some cases, it can fall outside this range in both directions). Benchmarking the models against numerical simulations, they found a relatively good agreement over a wide parameter range. 

In Figure \ref{fig:sfr_tff} we plot our estimate of the SFR against $M_{\rm cloud} / t_{\rm ff}$. The dashed lines in this plot are lines of constant $\epsilon$. We se that most of the  simulations lie between $\epsilon = 0.1$ and $\epsilon = 1$. One cloud, {\em n264-m156}, lies above the $\epsilon = 1$ line. Although the physics in our models differs from those of \citet{fk12}, both in terms of the way the turbulence is modelled and the thermodynamic treatment of the gas, we find that our clouds are in good agreement with their results.

Given that density appears in the derivation of the star formation rate per free-fall time, we would naively expect the SFR in our clouds to depend on the mean density of the cloud. In Figure \ref{sfrclouds} we plot how the star formation rate in the simulations varies with several of the cloud properties. In the upper left panel, we show how the SFR varies with the initial number density of the cloud. In this case, we have normalised the SFR with the cloud mass over the free-fall time, to be consistent with the definition in Equation \ref{eq:sfetff}. We see that there is not a significant trend in the data. Given that the $y-$axis spans roughly 1 order of magnitude while the $x-$axis spans 2 orders of magnitude, it would be tempting to say that the data is consistent with an $n^{1/2}$ scaling, and thus consistent with scaling as expected from the free-fall in the denominator. However, the scatter in the data is large, and if we were to remove only one or two outliers, the apparent trend would disappear. The observations reviewed in \citet{kdm2012} suggest that in real star-forming systems, $\epsilon$ cannot depend strongly on the mean density of the system, consistent with what we find here.

We have also explored how the SFR depends on the mass of the cloud. This is illustrated in the upper right panel of Figure \ref{sfrclouds}. There is some indication that the SFR increases as we move to more massive clouds, but we caution that again this apparent trend is very sensitive to the outliers of the distribution. If we were to remove the result of the {\em n2640-m20} run from the plot, the apparent trend would be much weaker. We also see that in massive clouds, there is a considerable amount of scatter, amounting to over an order of magnitude at the high mass end.

Since stars will form most easily in regions that have low Jeans masses, it is also interesting to examine whether there is a correlation between the SFR and the amount of cold, dense gas in the clouds. From our previous studies \citep[see e.g.][]{molina11,gc12c,clark12}, we know that there is a good correlation between CO and cold gas, since to form large quantities of CO, we need to shield the gas from the external radiation field, and this same shielding also strongly reduces the effects of photoelectric heating, allowing the gas to reach low temperatures. It is therefore natural to try to use CO as a measure of the cold, dense component of the clouds. Modelling the detailed $^{12}$CO and $^{13}$CO emission from our simulated clouds lies outside the scope of our present study, and so for this analysis we use a simpler measure of the amount of gas traced by CO. We define a quantity
\begin{equation}
M_{\rm CO} = \sum_i^{N_{\rm part}} \frac{m_i \, x_{{\rm CO}, i}}{X_{\rm C}},
\end{equation}
where $m_i$ and $x_{{\rm CO}, i}$ are the mass and CO abundance of particle $i$, $N_{\rm part}$ is the total number of SPH particles, and $X_{\rm C}$ is the elemental abundance of carbon, measured by number and with respect to the number of hydrogen nuclei. (In all of the simulations presented here, $X_{\rm C} = 1.4 \times 10^{-4}$). In the lower left-hand panel of Figure~\ref{sfrclouds}, we show how the SFR in our clouds varies as a function of $M_{\rm CO}$. We see from the Figure that the relationship between $M_{\rm CO}$ and the SFR is much tighter than the relationship between the SFR and total cloud mass, and has a slope of around 1. In this case, the density of the cloud does not introduce scatter: in low density clouds, only a small fraction of their total mass is traced by CO, while in higher density clouds, a larger fraction is traced by CO. This effect substantially tightens the relationship seen in the upper right-hand panel of Figure~\ref{sfrclouds}. 

If one were to divide both axes in the lower left-hand panel in Figure~\ref{sfrclouds} by a constant area, the resulting relationship between the surface density of star formation and the surface density of gas traced by CO would be linear, as suggested by \citet{big08}. The overall trend in our simulations is therefore consistent with both observational \citep{lada12}, and theoretical predictions \citep{klm11, kdm2012, phn2012} that the star formation rate correlates well with the mass in dense gas. Note, however, that these results do not say how the surface density of CO {\rm emission} in these clouds correlates with the SFR, and so one needs to be careful when comparing to the observed relation. In systems where the $^{12}$CO emission becomes highly optically thick, the gas that the emission traces will not be the cold, dense gas responsible for forming the stars, and so even if the mass of CO correlates well with the SFR, it is not clear that the same will be true for the CO luminosity.

We also find a good correlation if we consider the total area and total mass of our clouds (including the atomic envelope). This is shown in the lower right-hand panel of Figure \ref{sfrclouds}, in which we plot the surface density of the star formation rate in each cloud against its surface density (which remains at roughly the value set in the initial conditions over the timescales we consider here). The relationship between $\Sigma_{\rm SFR}$ and $\Sigma_{\rm gas}$ is steeper than linear, consistent with what is seen in the real ISM. However, we also see that there is considerable scatter in $\Sigma_{\rm SFR}$ in our simulations with high $\Sigma_{\rm gas}$, possibly explaining why observations that probe only a limited range in $\Sigma_{\rm gas}$ \citep[e.g.][]{heid10} generally fail to recover any clear correlation between $\Sigma_{\rm SFR}$ and $\Sigma_{\rm gas}$.

Finally, we note that the column and number densities derived in this section are not what would be derived observationally. Instead we have limited our analysis to the physical properties obtained from the models. Placing these results in the observational plane would require a detailed post-processing of the data that is significantly outside the scope of this paper.

\section{The A$_{\rm K}$ = 0.8 ``threshold'' for star formation}
\label{sec:ak08}

\begin{figure}
\centerline{
\includegraphics[width=3.3in]{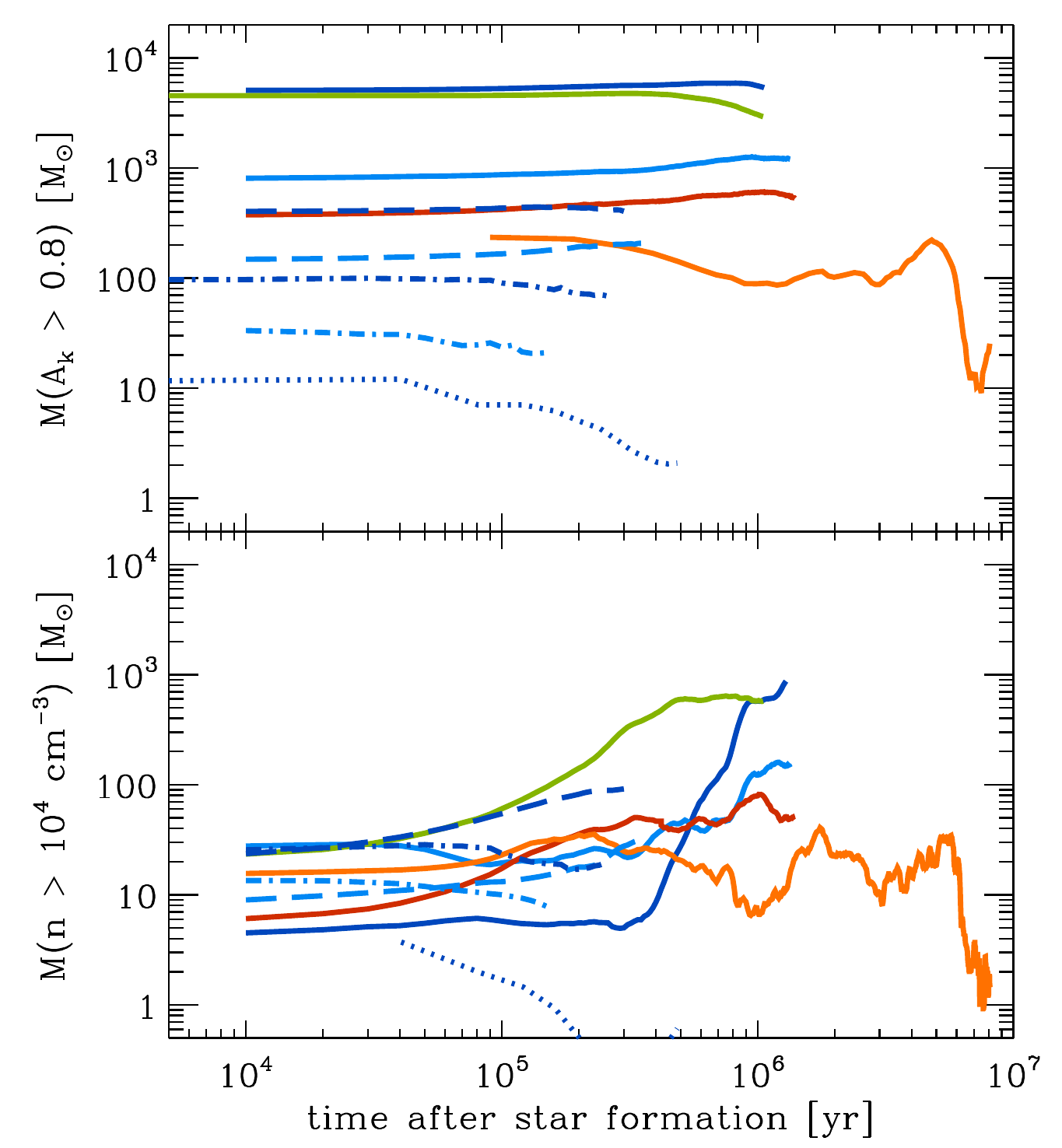}
}
\caption{\label{fig:mthresh} The top panel shows the mass residing above a column density of 0.025 g cm$^{-2}$, equivalent to a K-band extinction of 0.8, as a function of time after the onset of star formation in each cloud. In the bottom panel we show the mass above a number density of $10^4$ cm$^{-3}$, again as a function of time after the onset of star formation. All line-styles/colours are the same as those introduced in Figure \ref{fig:mstar_time}. The lines show that in our suite of clouds (which span a diverse range of conditions) the two mass fractions are not correlated. Further, we see that once star formation is underway, the mass above the column density threshold is approximately constant.}
\end{figure}

\begin{figure}
\centerline{\includegraphics[width=3.3in]{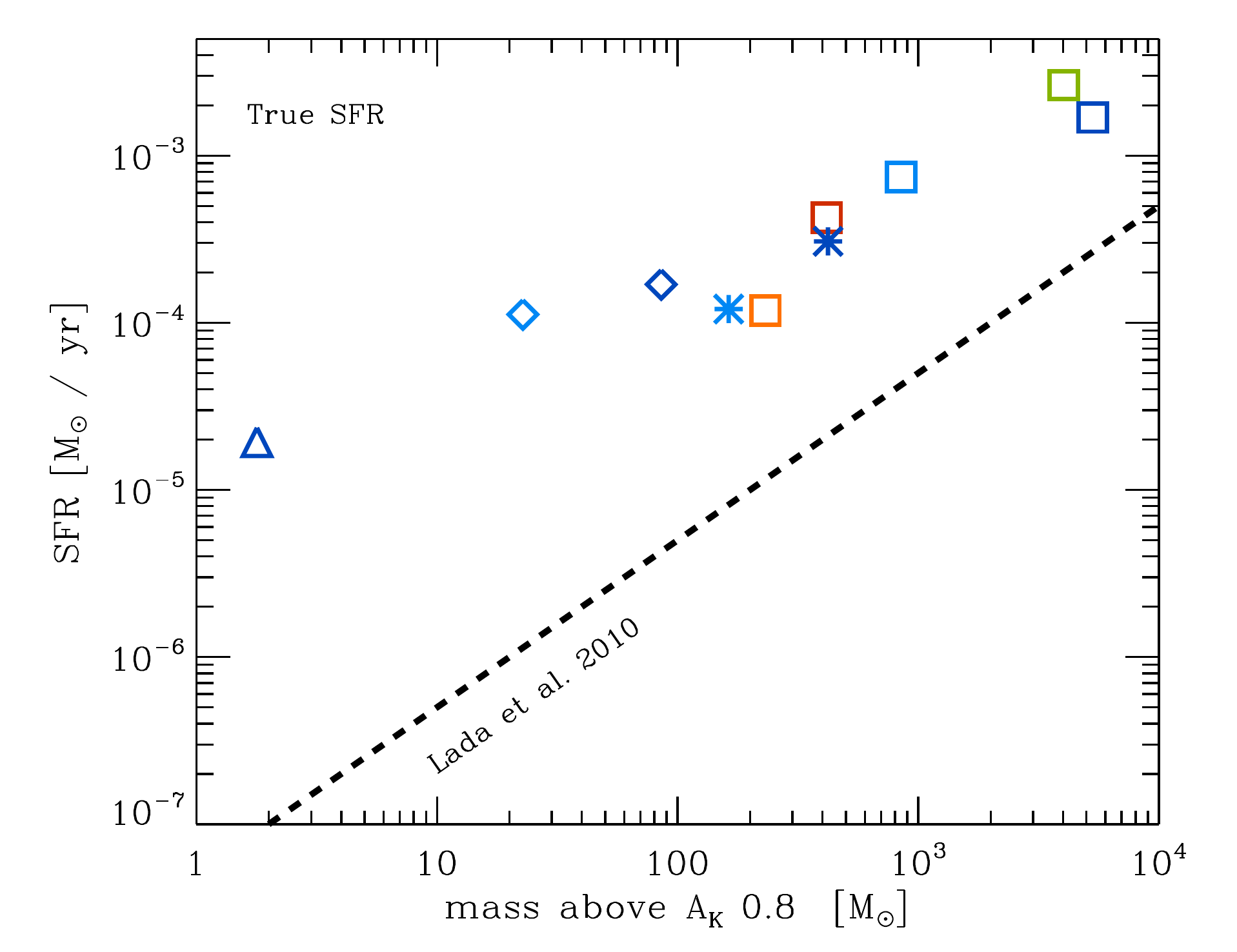}}
\centerline{\includegraphics[width=3.3in]{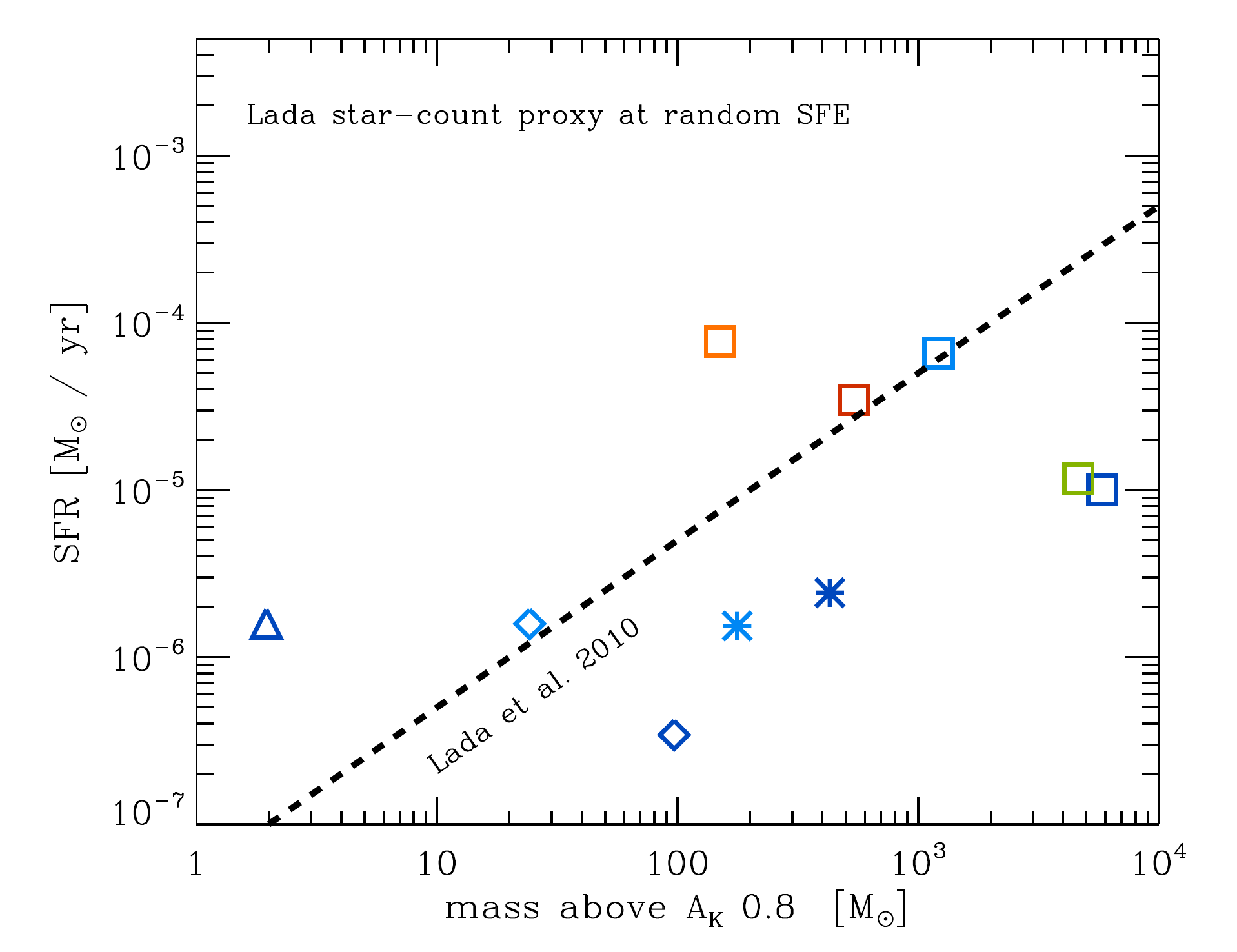}}
\centerline{\includegraphics[width=3.3in]{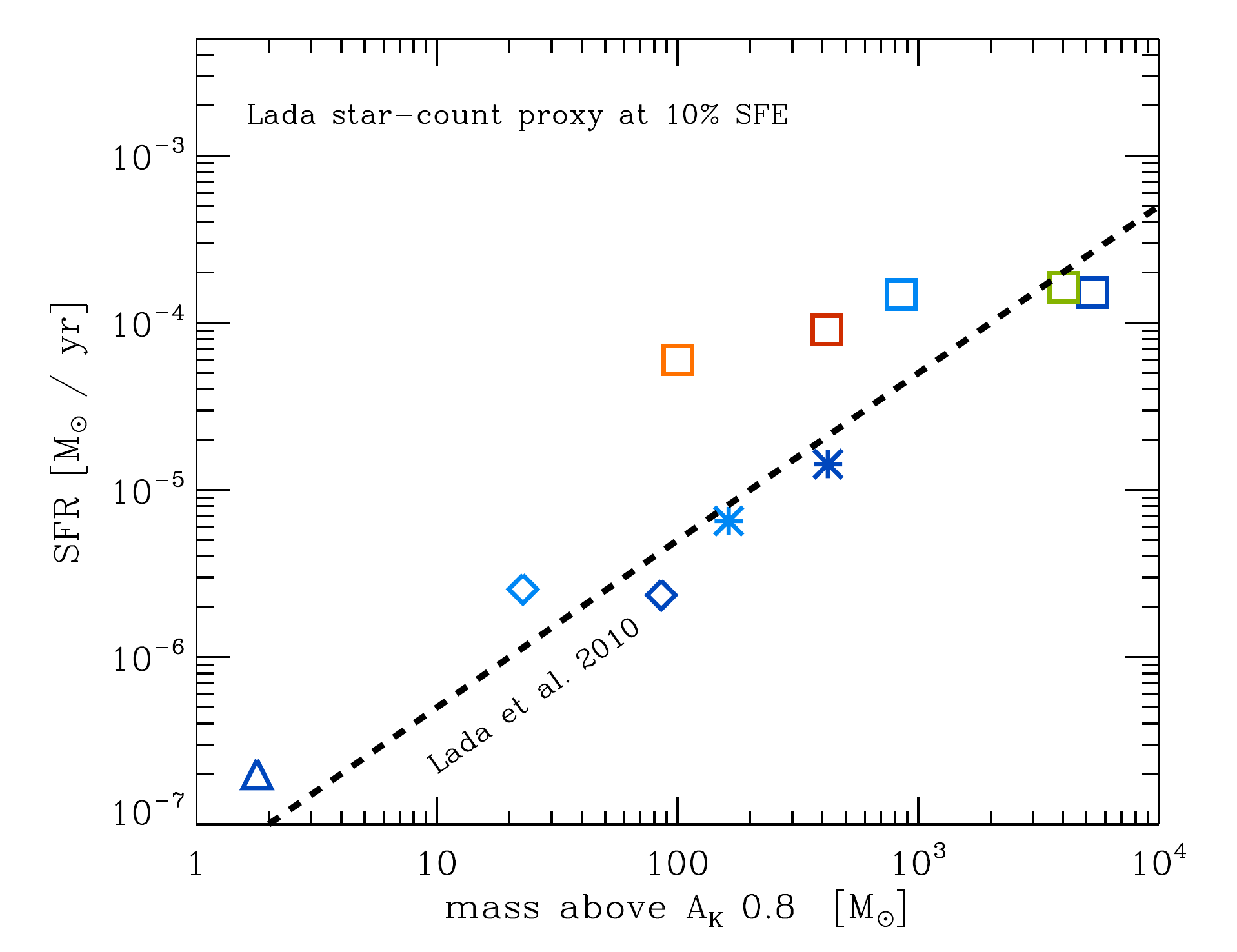}}
\caption{Star formation rate (SFR) as a function of the cloud mass located in regions with a line-of-sight K-band extinction $A_{\rm K} \geq 0.8$. In the top panel we plot the SFR obtained directly from the simulations by measuring the accretion rate onto the sink particles. In the middle and bottom panels, we attempt to reproduce the results of an observational-style YSO count to get the SFR. In the middle panel, the SFRs and column densities are drawn from random points during the star-forming period of the cloud evolution. In the bottom panel, the points are taken from the end of the simulations. See section \ref{sec:ak08} for more details. The black dashed line shows the relationship from the study of \citet{lada10}.}
\label{sfr_ak08}
\end{figure}

\citeauthor{lada10}~(2010; hereafter LLA10) showed that the number of young stellar objects (YSOs) in nearby molecular clouds is strongly correlated with the mass that resides above a K-band extinction of $A_{\rm K} = 0.8$, corresponding to a gas surface density of roughly 120 $\rm M_{\odot} \: pc^{-2}$. The relationship that they obtain can be written as
\begin{equation}
N_{\rm YSOs} = 0.2 \, \left(\frac{M_{K>0.8}}{\rm M_{\odot}}\right).
\end{equation}
If one assumes that the number of YSOs can be taken as a measure of the star formation rate, and further assumes a mean YSO age of 2 Myr and a mean YSO mass of 0.5 \solmasp, then their observations imply that the star formation rate in nearby clouds scales with $M_{K>0.8}$ as
\begin{equation}
{\rm SFR} \approx 5 \times 10^{-8} \, \left(\frac{M_{K>0.8}}{\rm M_{\odot}}\right) \, {\rm M_{\odot}} \, {\rm yr^{-1}}.
\end{equation}
The interpretation of this correlation put forward in LLA10 is that the mass above an extinction $A_{\rm K} = 0.8$ corresponds to the mass residing in regions with number density $n > 10^{4} \: {\rm cm^{-3}}$. As justification, they cite the fact that high density tracers such as N$_{2}$H$^{+}$ that are known to trace gas with $n > 10^{4} \: {\rm cm^{-3}}$ are typically observed only towards regions with $A_{\rm K} \sim 0.8$ or above \citep{bergin01,bergin02,aguti07}. Theoretically, we expect the amount of dense gas to correlate strongly with the star formation rate, so if we can identify this gas with the gas at $A_{\rm K} \geq 0.8$, then we will obtain a correlation between SFR and $M_{K>0.8}$, as observed.

\subsection{The connection between column and volume densities}

Although the interpretation of the data by LLA10 is seductive, one can see there is an obvious flaw in the reasoning: the fact that high density gas is found in high extinction regions does not necessarily imply that the high extinction regions are composed solely of high density gas. Surveys of Milky Way GMCs show that many have mean surface densities that are higher than 120 $\rm M_{\odot} \: pc^{-2}$ \citep[see e.g.][]{Solomon1987,rd10}, and yet the mean volume densities in these clouds are typically only a few times $100 \: {\rm cm^{-3}}$. Clearly the interpretation that all, or even most, gas residing at column densities greater than $120 \: {\rm M_{\odot}} \: {\rm pc^{-2}}$ has an intrinsic volume density above $10^{4}$ cm$^{-3}$ is overly simplistic, and only really applicable to isolated, starless cores, rather than the massive star-forming regions that make up the cloud sample in LLA10.  

We can use our simulations to test the assumption that the gas above the column density threshold corresponds to gas above number densities of $10^{4} \, \rm cm^{-3}$. In Figure \ref{fig:mthresh} we plot, for all the star-forming clouds in our study, the mass above a column density of 120 \solmas pc$^{-2}$ -- which we use as a proxy for $A_{\rm K} \geq 0.8$ -- as a function of time after the onset of star formation. The column densities were obtained from maps similar to those shown in Figure \ref{fig:cloudimages}. We also show the mass (in gas, not stars) residing above a number density of $n = 10^4 \,\rm cm^{-3}$.

The results show that the mass residing above the column density threshold is roughly constant once star formation is underway in our clouds. We also see that mass above the threshold is a function both of the mass and of the mean density of the cloud. Furthermore, these mass fractions are not particularly sensitive to our choice of random turbulent field: runs {\em n264-m10000} and {\em n264-m10000-s2} (the solid dark-blue and green lines), performed with different realisations of the turbulent field, yield very similar results for $M_{K>0.8}$. The only clouds that show a significant evolution in the mass above the threshold are {\it n5-m10000} (the solid orange line) and {\it n2640-m20} (the dotted, dark-blue line), both of which show a downward trend as they evolve. Interestingly, this is accompanied by a drop in the star formation rates of these clouds (see Figure \ref{fig:mstar_time}).

However, we see that the amount of gas above $10^4 \,\rm cm^{-3}$ in these clouds evolves quite differently. First, we see that the assumption that $A_{\rm K} \geq 0.8$ corresponds to gas with $n > 10^4 \,\rm cm^{-3}$ is a poor one, at least for these simple cloud models. At early times, the amount of gas above the threshold volume density can be orders of magnitude smaller than the amount of gas above the column density threshold, despite the fact that we are looking at period of the cloud evolution in which star formation is active. The second feature we note is that the mass above the volume density threshold is strongly time-dependent for most of the clouds in our suite. Although a cloud may contain a significant amount of mass above the column density threshold, not all of it is initially gravitationally bound. At first, only a single core (or small cluster-forming clump) goes into collapse, and so the mass above $n = 10^4 \,\rm cm^{-3}$ is around a few tens of solar masses. As progressively more of the cloud goes into collapse, and larger clusters are formed, the fraction of gas above $n = 10^4 \,\rm cm^{-3}$ rises. Gradually, the fractions of mass above the two thresholds do become quite similar, but only at fairly late times in the clouds' evolution, when our simulations are approaching a SFE of around 10 percent.

\subsection{The connection between A$_{\rm K} > 0.8$ and SFR in the models}
\label{sec:ourlla}
So what is the correlation between SFR and $M_{K>0.8}$ in our models, and do they reproduce the observed results? Given the limited resolution in our study, our sink particles are rather large -- roughly the size of a dense protostellar core -- and so we are unable to exactly reproduce the YSO count method used in LLA10 to estimate the SFR. However, we can create a mock YSO count by adopting some simple assumptions about the unresolved stellar population, that are motivated by previous studies. For example, it has been suggested that the efficiency in prestellar cores (on scales similar to those captured by our sink particles) is around 30-50 percent \citep{mm2000, alves2007, Machida2009, Machida2012, Price2012}. The total mass in the emerging YSOs is therefore the fraction, $\epsilon_{\rm core}$, of the total mass that ends up in cores, which we assume here to be given by $M_{\rm sinks}$, the total mass in sink particles. Finally, we can get an estimate of the number of YSOs by simply dividing by a mean stellar mass, $\bar{m}_{\rm star}$. Together, we get the following expression for estimating the total number of YSOs ($N_{\rm YSO}$),
\begin{equation}
\label{eq:nyso}
N_{\rm YSO} = \frac{M_{\rm sinks} ~ \epsilon_{\rm core}} {\bar{m}_{\rm star}}.
\end{equation}
Following LLA10, we take $\bar{m}_{\rm star} = 0.5 \,\rm M_{\odot}$, and adopt the above-quoted value of $\epsilon_{\rm core} = 0.3$. Note that while our choices of $\bar{m}_{\rm star}$ and $\epsilon_{\rm core}$ affect the normalisation of our results, they will not affect the slope.

An assumption of the LLA10 analysis is that the mean YSO mass is 0.5 $\rm M_{\odot}$. While this is a reasonable assumption for the more active star-forming regions in the sample, it likely overestimates the mean mass in lower-mass, or less active, regions, simply because the IMF in such regions is poorly sampled \citep{McKeeOffner2010}.  As a result, the star formation rates derived from the YSO counts in LLA10 are likely to be a factor of a few too high for these regions. For consistency with the analysis in LLA10, we use a mean mass of 0.5 $\rm M_{\odot}$ for our analysis. However, the reader should keep this caveat in mind when interpreting the results of YSO-count derived SFRs.

Another caveat of the YSO count method is that it only applies for objects between Class I and Class III. LLA10 adopt a mean age of 2 $\pm 1$ Myr for these objects, with the implication that YSOs will evolve beyond Class III after this time. For the majority of the simulations that we present here, this is not a problem: almost all the simulations are stopped before 2 Myr, and so our assumption that the mass we see is in the form of objects younger than Class III is a good one. However, three of the clouds may present problems. The first is cloud {\em n5-m10000} (orange solid line), since it accretes for nearly 8 Myr. As such, we limit the analysis of this cloud to the conditions at $< 2$ Myr after the onset of star formation. In contrast, clouds {\em n264-m10000} and {\em n264-m10000-s2} are undergoing a burst of star formation at late times, and so it is likely that many of the sink particles in these simulations represent Class 0 objects. However, Figure \ref{fig:mstar_time} shows that the total mass in sink particles varies by less than a factor of 2 over timescales of around a few times $10^5$ yr, consistent with the expected lifetime of this phase \citep{evans09}. As such, the effect is small, given the other errors, and we simply caution the reader that the YSO count obtained from Equation \ref{eq:nyso} is an over-estimate. 

The results from this analysis are presented in Figure \ref{sfr_ak08}. In the top panel, we show the SFR obtained by averaging the sink particle accretion rates over the star-forming period of the simulation. These are the same as the accretion rates presented in Section \ref{sec:sfr}, and since they are obtained directly from the simulations we refer to them as the ``measured sink SFR'' here. In the middle and bottom panels, we use the SFR as defined via the YSO count method. In the middle panel, we sample the SFR and column densities at random points in the evolution of each cloud (note that for cloud {\em n5-m10000} this is done only at times within 2 Myr after the onset of star formation). In the bottom panel, we sample the SFR at and column densities at the end of the simulation (or 2 Myr after $t_{\rm SF}$ in the case of cloud {\em n5-m10000}).

We see that the measured sink SFR lies significantly above the relation presented by LLA10, by at least an order of magnitude. The intrinsic properties of our simulations are therefore not consistent with the observational picture -- star formation occurs too quickly in the high density gas. While the situation is worse for the low-mass clouds, this is to be expected. These clouds have only a few accreting sink particles at any time during their star-forming period, and due to the $\sim 10$\,K gas that creates the star-forming cores, their accretion rate is expected to be between 10$^{-6}$ and  10$^{-5}\, \rm M_{\odot} yr^{-1}$ \citep{Shu1977}. As such, the minimum accretion rate that these clouds can achieve is limited by the thermodynamics of the dust. One obvious reason for the large discrepancy between the observations and our measured sink SFR is that in the simulations, we assume that all of the mass accreted by our sink particles ends up in stars. Because of the large size of our sinks, this is essentially equivalent to assuming that $\epsilon_{\rm core} = 1.0$. If we instead set $\epsilon_{\rm core} = 0.3$, as in our YSO count model, then our derived SFRs will systematically decrease by around a factor of three. They will still lie above the observed correlation, but the discrepancy is much less pronounced once this effect is considered, with the simulations now lying only around a factor of two above the observed relation. The remaining discrepancy may be a consequence of our neglect of magnetic fields in these simulations, as the additional support that they provide can easily reduce the SFR by an additional factor of a few  \citep[see e.g.][]{PriceBate2008, wang2010}.

In the middle panel in Figure \ref{sfr_ak08}, we use our mock YSO count to model the SFR in an attempt to get closer to what the observations are actually measuring. We see that this crude attempt to model the observational analysis results in a much better agreement with the observed relation. However the scatter is large, with points lying both around an order of magnitude above and below the LLA10 relation.

In the bottom panel, we plot the SFR derived from our mock YSO counts using information from the simulations at the point when they have converted $\sim 10$ percent of their mass into sink particles (with the exception of cloud {\em n5-m10000}, as stated above). Now we see a much better agreement than we had in the middle panel, with most of the points lying within a factor of a few from the observed relationship. The main outlier here is our low density cloud {\em n5-m10000}, which still has a very high SFR for its given mass above the column density threshold. Given that this cloud is collapsing from extremely low densities, it is perhaps unsurprising that it does not lie on the relation with the rest of the clouds.

\begin{figure}
\includegraphics[width=3.3in]{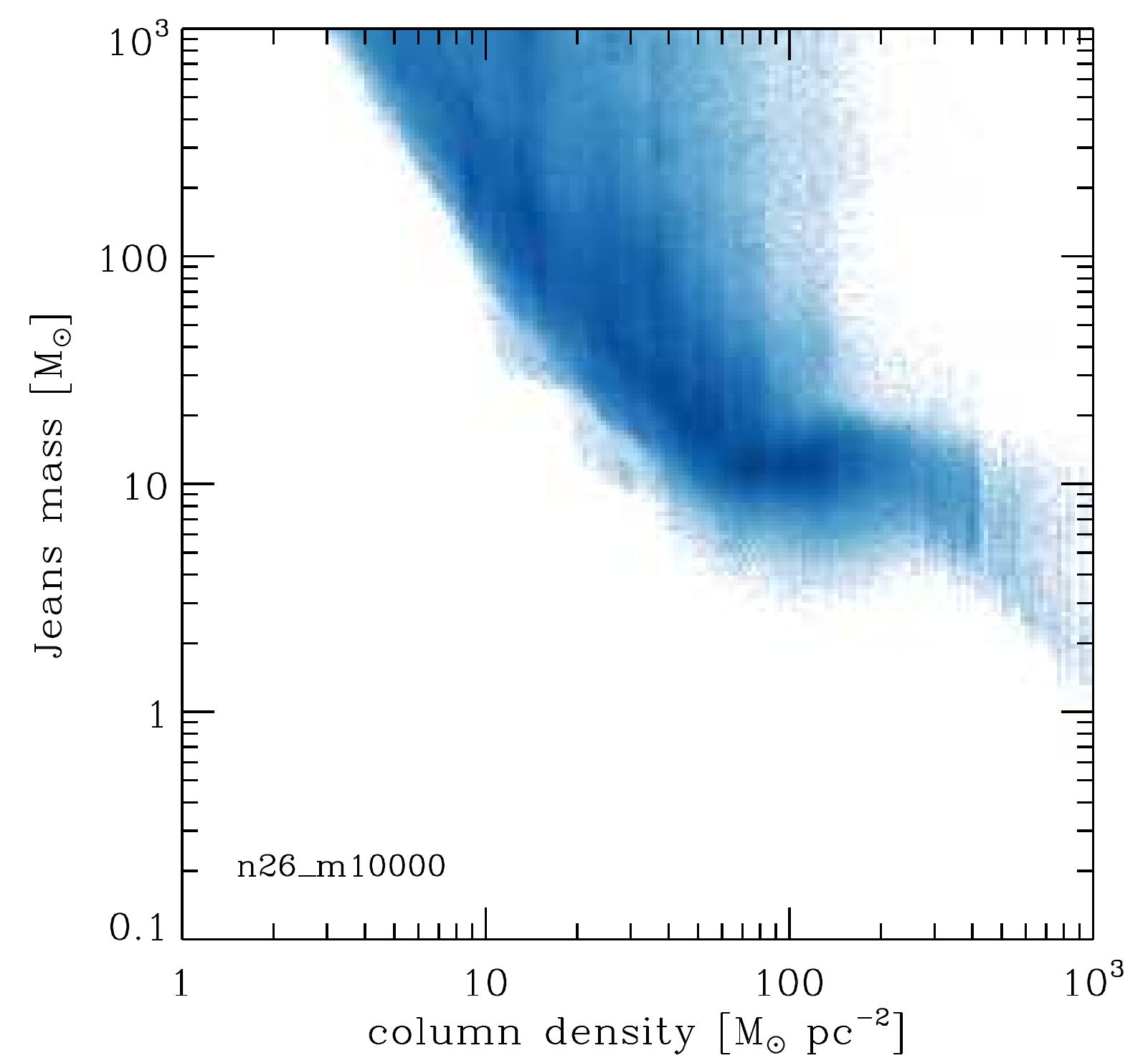}
\includegraphics[width=3.3in]{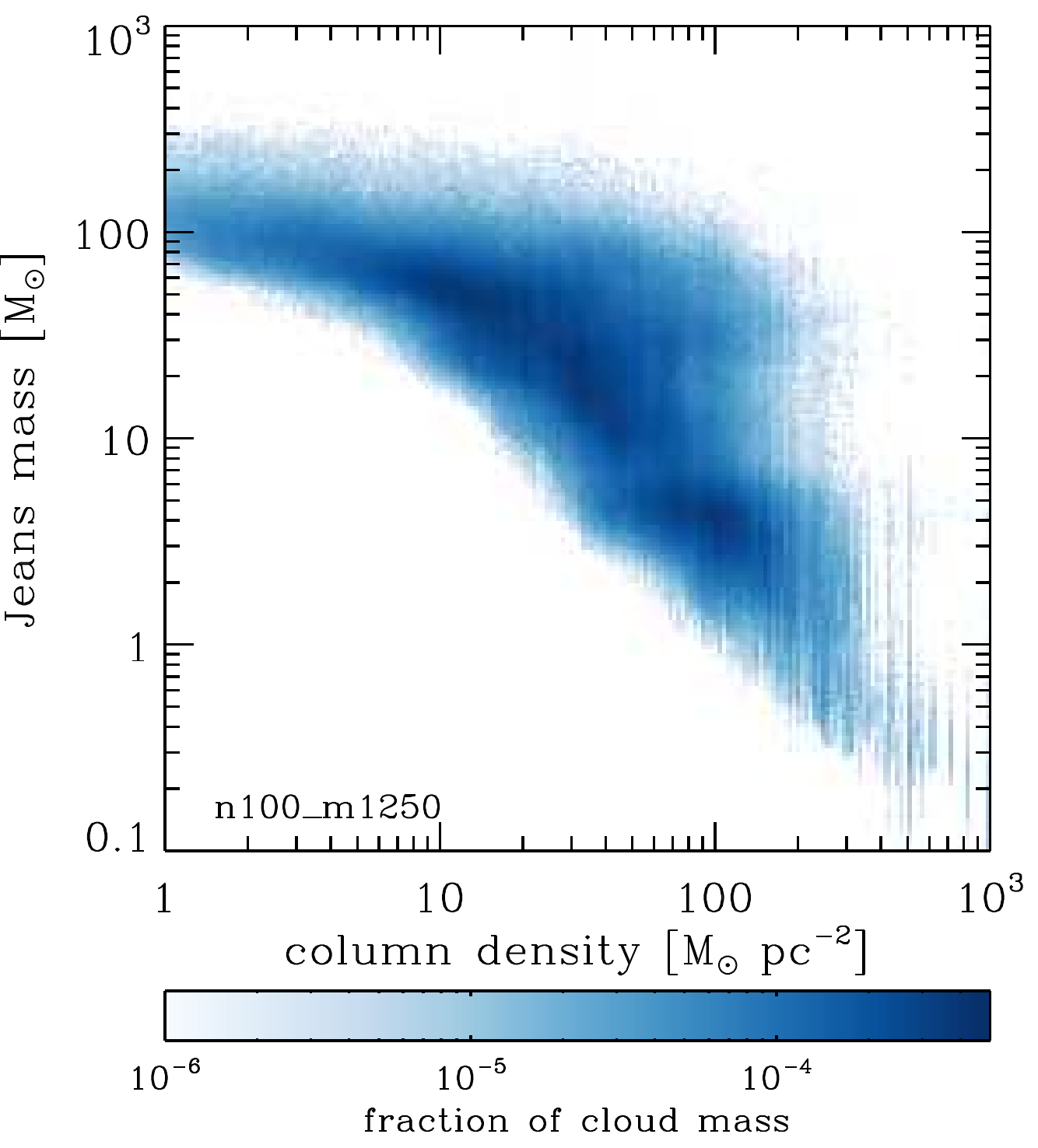}
\caption{Variation of the particles' Jeans mass (calculated from the particles' temperature and density) with the projected column density, as seen looking down the $z$-axis in the simulation. The column density images used here have a resolution of 0.02 pc. We see that along a given line of sight, the Jeans mass can vary substantially. However, for regions with a column density above $~$100 \solmas pc$^{-2}$, the scatter drops substantially. This reflects the column density at which the gas can shield itself from the ISRF, thereby reducing the effect of the photoelectric emission heating.}
\label{fig:av_jeans}
\end{figure}

\section{The origin of the A$_{\rm K} > 0.8$ ``threshold''}
\label{sec:ak08-origin}

So what is the origin of the ``threshold'' column density? Is it simply that the dense gas fraction (eventually) matches the fraction of gas above the column density threshold -- as discussed above -- or is there something else going on? We can gain considerable insight into the answer to this question by examining how the local Jeans mass of the gas
varies as a function of the line-of-sight column density. This is illustrated in Figure~\ref{fig:av_jeans} for two of our model clouds. To construct this Figure, we first calculated the Jeans mass for each SPH particle, using its local values of the density, temperature and mean molecular weight. We then binned the particles by their $(x,y)$ position, using a grid with a cell size of 0.02~pc, and calculated the column density of the gas in each bin. Finally, we assumed that each particle in a bin shared the same projected column density (i.e.\ we ignored variations on scales smaller than the cell size of the grid). The vertical striations we see in Figure~\ref{fig:av_jeans} are therefore simply due to the fact that many SPH particles -- with a range of densities and temperatures -- contribute to one point in the column density image. 

For both of the clouds shown here, we see the same general trend in the Jeans mass distribution: between 10 to 100 \solmas pc$^{-2}$ the scatter in the Jeans mass is over two orders of magnitude, but at higher columns, the scatter abruptly decreases. The reason for the sudden decrease in the scatter is that once the projected column density reaches
a value of around 100 \solmas pc$^{-2}$, the gas becomes shielded from the ISRF to the point where the heating effects from the photoelectric emission become negligible. 

\begin{figure}
\centerline{\includegraphics[width=3.4in]{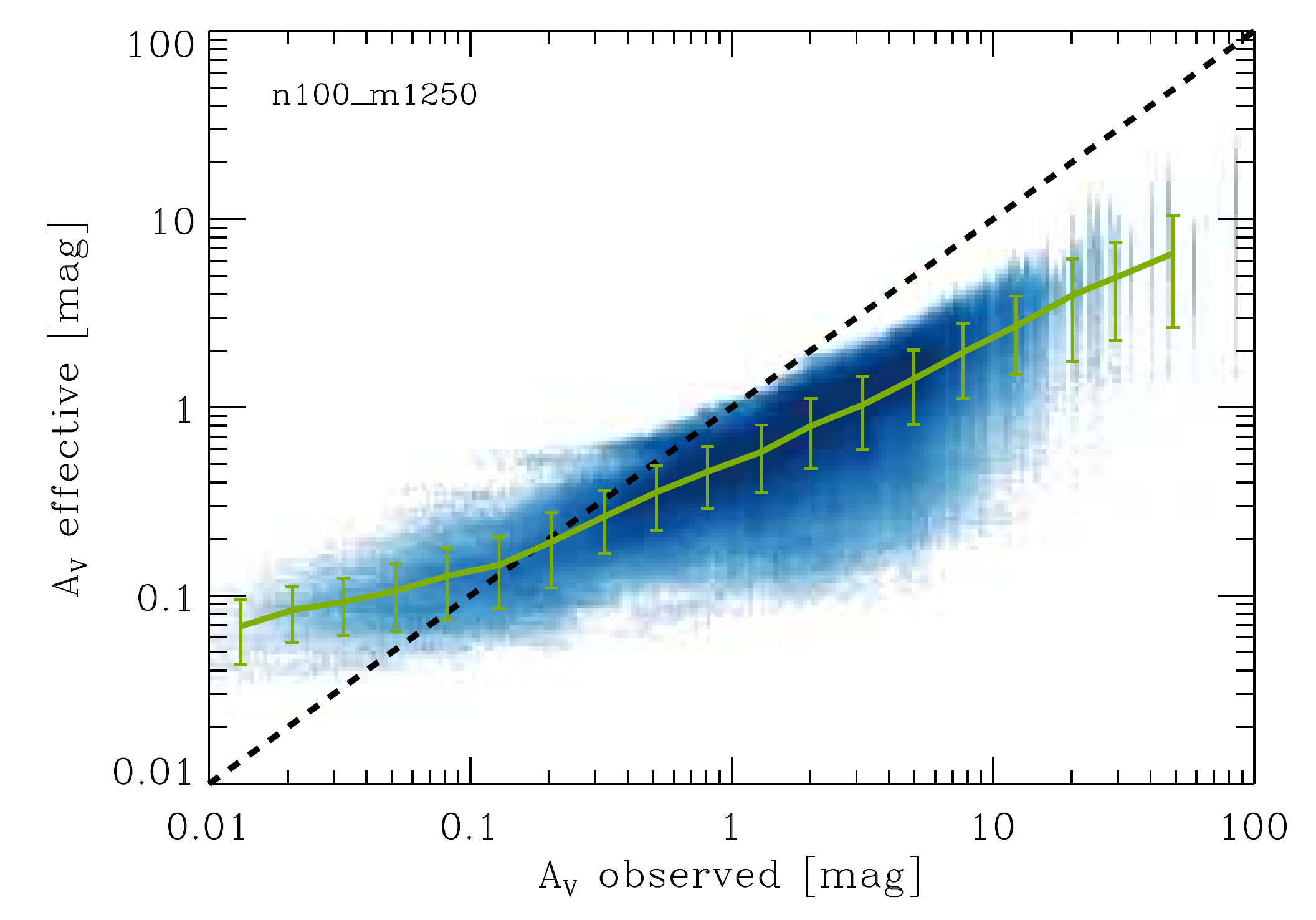}}
\centerline{\includegraphics[width=3.4in]{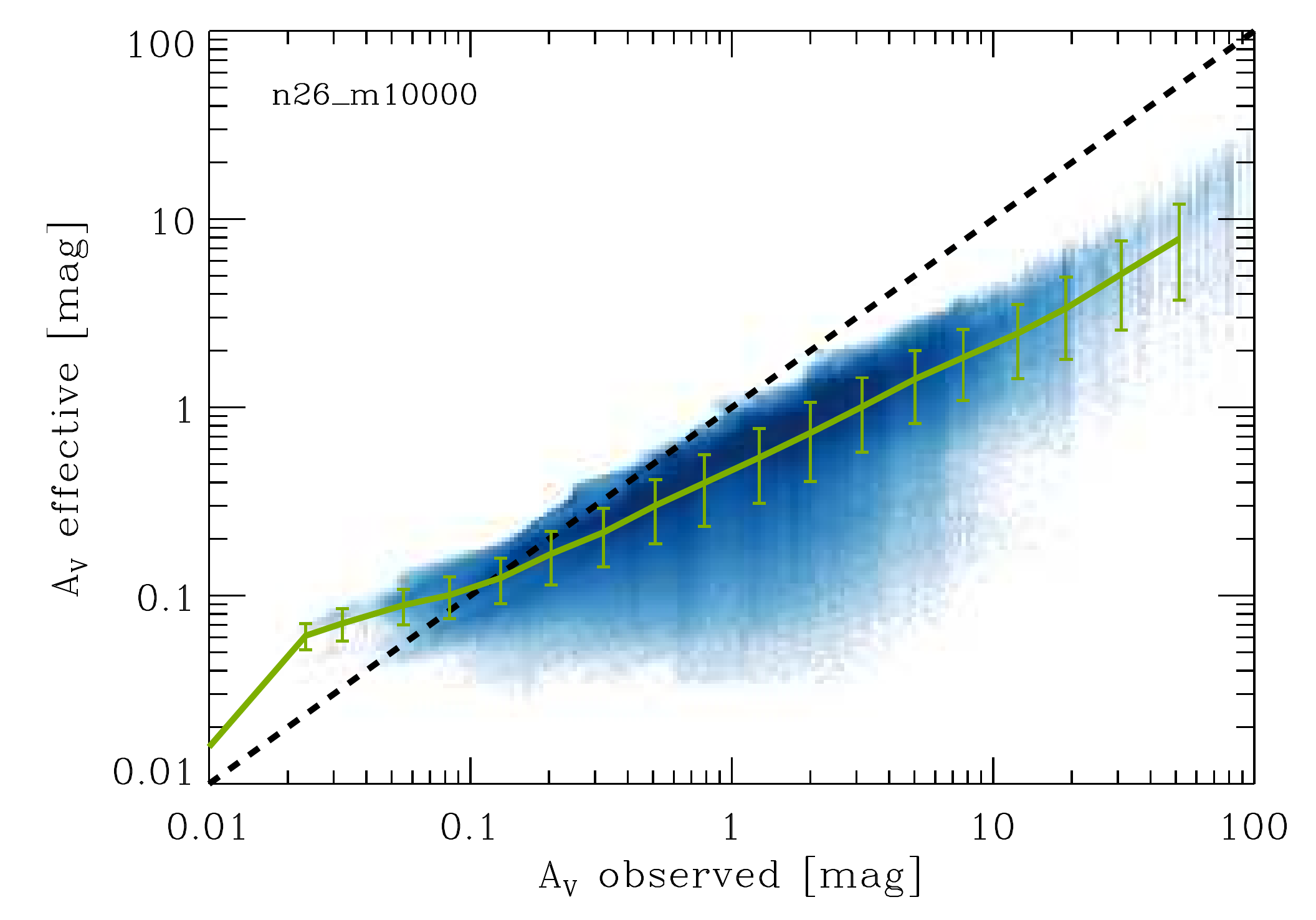}}
\centerline{\includegraphics[width=3.4in]{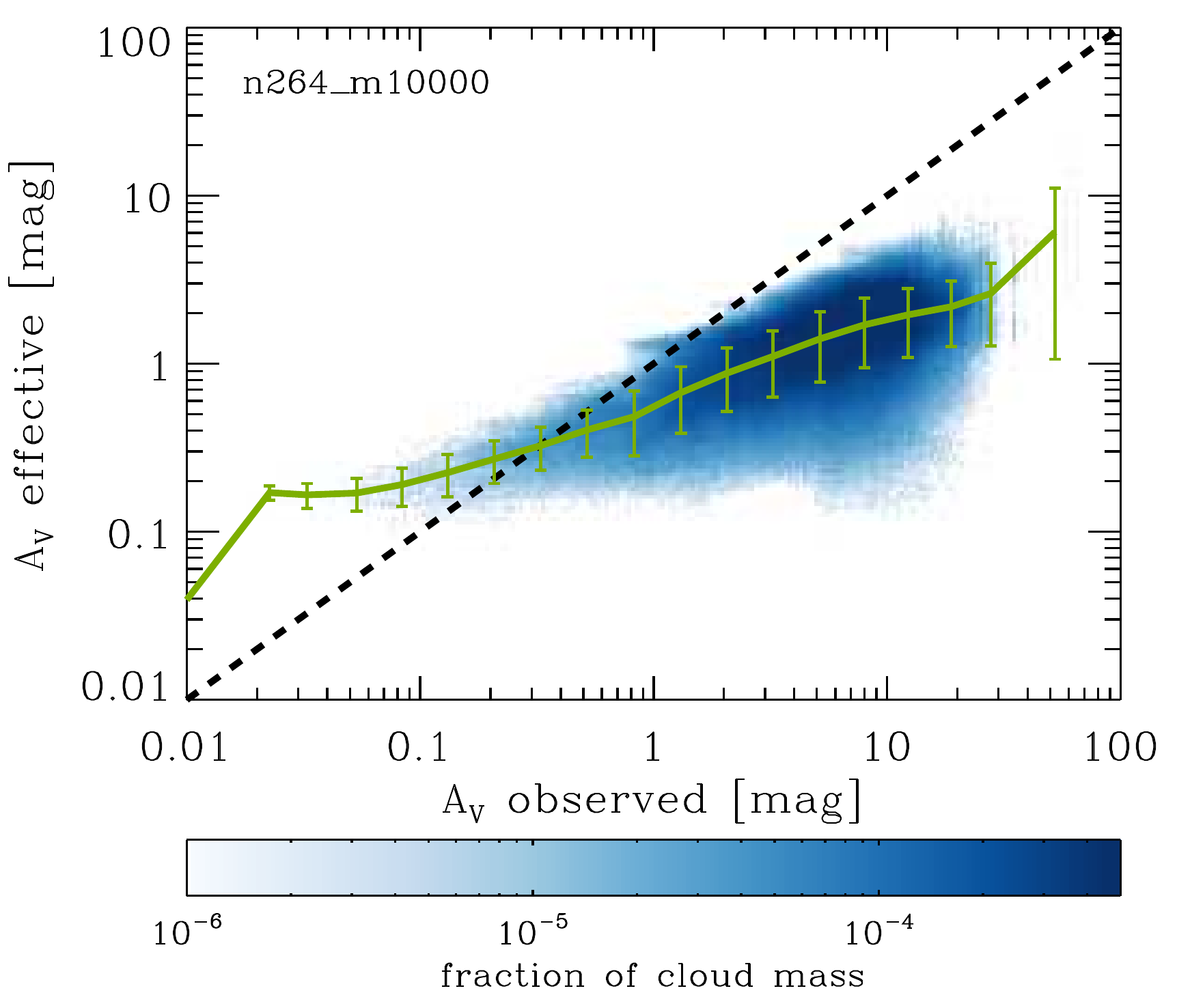}}
\caption{We compare the ``effective'' visual extinction, $A_{\rm V, eff}$, for each SPH particle -- that is, the angular averaged $A_{\rm V}$ as seen by the SPH particle during the simulation -- with the particle's ``observed'' extinction, $A_{\rm V, obs}$, as would be derived from our line-of-sight column density maps (such as those in Figure \ref{fig:cloudimages}). The distributions are shown for three contrasting cloud models. Our values of $A_{\rm V, eff}$ are calculated via Equation \ref{eq:aveff}, and once again we use the data from immediately before the onset of star formation in these clouds. Green lines and error bars show the mean and standard deviation of the distribution.}
\label{fig:aveff}
\end{figure}

A projected column density of around 100 \solmas pc$^{-2}$ corresponds to a visual extinction $A_{\rm V} \sim 5$, significantly higher than the value of $A_{\rm V} \sim 1$ at which photoelectric heating is generally thought to stop playing a central role in regulating the gas temperature in molecular clouds. There are two reasons for this apparent discrepancy. First, the projected column density measured along a particular line of sight includes all of the material along that line of sight from one boundary of the cloud to the other. However, even in a uniform cloud, the column density between any point on that line of sight and the nearest cloud boundary can never be larger than half this value. Second, the turbulence in the cloud opens channels through which the radiation can penetrate \citep[see e.g.][]{bethell07}, and hence the angle-averaged value of the column density about a given point on the line of sight will generally not be the same as the value measured along the line of sight. For points along high column density lines of sight, the angle-averaged or ``effective'' extinction, $A_{\rm V, eff}$, will typically be lower than the line of sight value, while the opposite is true for the lowest column density sight-lines.

This effect is illustrated in Figure~\ref{fig:aveff}, where we show plots for several of our simulations that compare the effective visual extinction for each SPH particle with the ``observed'' extinction, $A_{\rm V, obs}$, i.e.\ the value measured along the line of sight passing through the SPH particle. To compute the effective visual extinction, we use the following formula
\begin{equation}
\label{eq:aveff}
A_{\rm V, eff} = - \frac{1}{2.5} \ln \left[ \frac{1}{N_{\rm pix}} \sum_{i = 1}^{N_{\rm pix}}
\exp \left(-2.5 A_{{\rm V}, i} \right) \right],
\end{equation}
where $A_{{\rm V}, i}$ is the extinction corresponding to the column density in the direction defined by pixel number $i$, and we sum over all $N_{\rm pix}$ pixels in our {\sc TreeCol} column density map.\footnote{The weighting factor of 2.5 that we use here is motivated by the fact that the photoelectric heating rate scales with $A_{\rm V}$ as  $e^{-2.5 A_{\rm V}}$.} 

We see from Figure~\ref{fig:aveff} that most of the mass along lines of sight with $A_{\rm V, obs} < 5$, corresponding to a projected column density $< 100 \, {\rm M_{\odot} \, pc^{-2}}$, has a low effective extinction and hence will be strongly heated by the ISRF. Along lines of sight with $A_{\rm V, obs} > 5$, on the other hand, we find that typically $A_{\rm V, eff} > 1$--2, implying that photoelectric heating will be unimportant in the majority of this gas. 

Given these results, we propose that the observed column density threshold reflects the extinction required for clouds to shield themselves from their environment. It is not a ``threshold'' in the precise sense of the word, but rather a marker -- above this column density, the observed amount of mass and the number of Jeans masses are well correlated, while at lower column densities there is considerable scatter. Some support for this idea comes from observations that find a steep dependence of $\Sigma_{\rm SFR}$ on $\Sigma_{\rm gas}$, rather than a sharp threshold (see e.g.\ \citealt{guter11} or the recent review of \citealt{molinari14}). Note that although the initial conditions for the two clouds represented in Figure \ref{fig:av_jeans} are very different, the average (mass-weighted) Jeans mass around the threshold column density varies by no more than a factor of 2--3 from one cloud to the other. This also suggests that above the threshold column density, the conditions in the star-forming regions of the cloud tend to be similar, regardless of the more general conditions of the parent cloud, although there is a hint that the Jeans mass in the shielded gas may increase somewhat with increasing cloud mass. What effect this has, if any, on the mass function of the stars formed in the cloud is beyond the scope of this paper. Further, the fact that column density enters the photoelectric heating term via an exponential suggests that variations of even a factor of 10 in the strength of the local ISRF should not produce much variation in the threshold column density, although much larger changes may have a more pronounced effect. 

Finally, we should note that an alternative explanation for the observations is simply that the A$_{K > 0.8}$ threshold draws a boundary that is less polluted by line-of-sight clouds that are nothing to do with the star formation complex in general. Our current models are unable to test this hypothesis, but a suite of models that account for the large-scale flows in which clouds form (e.g. \citealt{enrique2006,Heitsch2006,vs07,Hennebelle2008,banerjee09,enrique2009,rg10,clark12}) may be able to address this question in the future.

\section{Discussion}
\label{discuss}
The results of Section \ref{sec:ourlla} show that measured sink SFR in the simulations, lie significantly off the relation observed by LLA10 for nearby clouds. However, when we attempt (admittedly crudely), to reproduce the SFR estimate based on YSO counts, as used by LLA10, we find that our simulations are in much better agreement with the observations, especially at late times (Figure \ref{sfr_ak08}).

If one assumes that the simulations are a good representation of the evolution of local clouds, then the implication is that the YSO count method is systematically underestimating the true SFR in these clouds. However it may not be by much: the points in the top panel of Figure \ref{sfr_ak08} represent an upper limit on the actual star formation rate, since they assume that all of the mass accreted by sink particles is turned into stars. As our sink particles are better thought of as representing individual pre-stellar cores rather than individual stars, it is not at all clear that the actual star formation efficiency will be so high. Adopting a standard star formation efficiency factor for the gas in the sink particles (or prestellar cores) of $\epsilon_{\rm sink} = 0.3$ brings the results for the higher-mass clouds in the top panel of Figure  \ref{sfr_ak08} down to within a factor of 2 of the LLA10 observations. (As discussed in Section \ref{sec:ourlla}, we do not expect the lower mass clouds to fall on the line). Given that the errors on the YSO method are around a factor of 2 \citep{evans09}, the simulations and observations could then be in agreement. It should also be noted that calibration of the YSO count method implicitly assumes that the SFR in the cloud was constant in time. We can see from Figure \ref{fig:mstar_time} that this is clearly not the case, and that most of the clouds experience some sort of acceleration in their SFR. Such an acceleration in real clouds would result in an overpopulation of one particular class of protostellar object, in comparison to the assumption of constant SFR, and would thus result in an overestimate of the class lifetime.

That said, the numerical simulations presented here are clearly not without their shortcomings. Currently, the simulations do not include magnetic fields. These have been shown to reduce the rate of star formation by a factor of a few, which would again bring the models into better agreement with the observed SFR relations \citep[e.g.][]{PriceBate2008, wang2010}. In addition, it is likely that feedback in the form of winds and radiation from higher mass stars would also lower the overall rate of star formation in the clouds \citep[e.g.][]{Cunningham2011, Dale2012, kkm2012, Dale2013}. However, until the numerical models can capture all the physics required to create realistic YSO emission -- and recent advances have been made towards this end \citep{offner11, offner12} -- it is impossible to determine whether the real YSO counts are in fact systematically underestimating the true SFR in nearby star-forming regions; the work we present here merely suggests that such a follow-up study would be worthwhile.

Our finding in Section \ref{sec:ak08} that the mass at high column densities -- above 120 \solmas pc$^{-2}$ -- is not well correlated to the mass at densities above $10^4 \rm cm^{-3}$, has an implication for the ``slow star formation'' model as presented by \citet{Tan2006}. Lada et al. (2010) propose that the star formation rate is modelled by (their Equation 3),
\begin{equation}
{\rm SFR} = \epsilon \, \frac{M_{n>10^4\,\rm cm^{-3}}}{t_{\rm ff}(10^4\,\rm cm^{-3})},
\end{equation}
were $\epsilon$ is the star formation rate per free-fall time (as discussed in Section \ref{sec:sfrprops}), $t_{\rm ff}(10^4\,\rm cm^{-3})$ is the free-fall time of the gas at $10^{4} \,\rm cm^{-3}$, and $M_{n>10^4\,\rm cm^{-3}}$ is the mass residing at number densities greater than $10^4 \,\rm cm^{-3}$, which LLA10 assume is equivalent to the mass in gas above the column density threshold of A$_{\rm K}$ = 0.8. In order to get good agreement with their observational relation between the SFR and the mass above the column density threshold, they require that $\epsilon = 0.018$. In these circumstances, the star formation would be said to be slow, since very little of the ``high density'' gas is being converted to stars in a free-fall time. However, we have seen in this study that the correlation between gas at high column densities and the gas at high volume densities can be poor, even though the clouds still follow the SFR relation proposed by LLA10. In our simulations, the star formation rate per free-fall time is actually greater than 0.1 at densities of around $10^4 \,\rm cm^{-3}$ and above. Thus if $M_{n>10^4\,\rm cm^{-3}} \neq M_{A_{\rm K> 0.8}}$ in real star-forming clouds -- and from our discussion in the previous section, it would seem that this is likely -- then the claims made by LLA that their analysis is in support of \citet{kt07} are unfounded. 

However, \citet{kdm2012} have shown that the star formation rate per free-fall time based on the LLA10 and \citet{heid10} data is consistent with a star formation rate per free-fall time of 0.01 to 0.03, similar to the estimates from \citet{kt07} for high-density gas, which are based on the HCN emission \citep{gs04a, gs04b, wu05}. As such, it seems that there is still a discrepancy between the observational SFR estimates, and those found in the simulations, such as our calculations presented here, as well as those of other groups \citep{Bate2012, kkm2012, fk12}.

Another implication of our results is that LLA10's relationship is not an {\rm evolutionary} process, but rather that clouds are born with an inherent ability to form stars. In our clouds, the mass above the column density threshold remains roughly constant once star formation has set in, but we know from Figure \ref{fig:mstar_time} that the mass in stars is constantly rising after this point. We have seen that if the clouds are caught at random  stages of their evolution, then the correlation between $N_{\rm YSOs}$ (or the mass in stars) and  $M_{K>0.8}$ is relatively weak (middle panel of Figure \ref{sfr_ak08}). The fact the true clouds exhibit a fairly tight correlation, such as we find in the bottom panel of Figure \ref{sfr_ak08}, implies that we are catching them at a similar stage in their star formation cycle, again assuming that the clouds in our study are representative of those studied by LLA10. Indeed, LLA10 adopt a mean value of $2 \pm 1$ Myr for the ages of their YSOs, to account for the wide range of protostellar ages in their cloud sample. The implication is that these regions have been forming stars for at least $2$ Myr, consistent with the fact that there are Class III sources present in the clouds. 

\section{Conclusions}
\label{conc}

We present a suite of turbulent cloud calculations that span a wide range in mass and radius, and thus sample a range of possible volume and column densities. Our goal is to determine whether there exists a threshold -- in either volume or column density -- below which star formation is suppressed. The calculations were performed using a modified version of the publicly available SPH code {\sc GADGET-2} (Springel 2005), which includes a sink particle implementation to handle regions of star formation, and a time-dependent chemical network that is coupled to a model of the ISM that accounts for the main heating and cooling processes. In addition to these 3D SPH calculations, we perform a large number of one-zone cloud models that give a more complete coverage of the parameter space. We find that the one-zone models are generally a good indicator of star-forming ability of the full 3D SPH models.

The results of our study can be summarised as follows:
\begin{itemize}

\item We find that there exists a mean column density of $\sim 10^{21}\,$cm$^{-2}$ (hydrogen nuclei), below which clouds, or regions within clouds, are unable to form stars. These results demonstrate that clouds with properties similar to the general ISM in spiral arms of the Milky Way are capable of forming stars, and that the $\sim 116$ \solmas pc$^{-2}$ ($10^{22} \,\rm cm^{-2}$) ``threshold'' for star formation, as reported by \citet{heid10} and \citet{lada10} is simply a consequence of the star formation process, rather than a prerequisite for it.

\item We find that for clouds with masses of the order 1000 \solmas and below, there exists a cloud-averaged column density threshold of $\sim 10^{21}\,$cm$^{-2}$, below which clouds are sterile. Clouds with larger masses ($10^4$ \solmasp) are still able to form stars down to $\sim 5 \times 10^{20}\,$cm$^{-2}$, with resulting volume densities of around $5\,$cm$^{-3}$. However, the bound, collapsing sections of the cloud have a mean column density in excess of $\sim 5 \times 10^{21}\,$cm$^{-2}$, suggesting that this column density does represent some sort of threshold below which star formation (cloud collapse) does not occur. We find that the ability of a cloud to form stars is well described by a simple Jeans mass argument. 

\item We find that the temperature and density PDFs in the sterile clouds are narrower than those in the star-forming clouds. In addition, the density PDFs of the star-forming clouds display the same power-law profile at high densities that has been reported elsewhere in the literature.

\item The mass above a column density of $\sim 120$ \solmas pc$^{-2}$ correlates well with the star formation rate (SFR) in our clouds, when we adopt a YSO count style measure of the SFR, similar to that employed by \citet{lada10}. As discussed above, we find no evidence that $120$ \solmas pc$^{-2}$ is a ``threshold'' for star formation. Instead we suggest that this column density traces regions in which the gas is able to shield itself from the ISRF and the heating effects of photoelectric emission, consistent with previous studies \citep{schaye04,klm11,gc12a}. If cold gas is found predominantly along lines of sight with column densities above this ``threshold'', and stars form mainly in cold, dense regions, then the result is a relationship between the amount of gas above the threshold and the star formation rate, as observed. Note also that as stars form in over-dense regions within the clouds, it is unsurprising that the column density of the star-forming regions is significantly higher than the cloud-averaged column density threshold for star formation discussed above.

\item Our simulations show an extremely poor correlation between the mass with volume densities $n > 10^4 \,\rm cm^{-3}$ and the mass residing at column densities of $120$ \solmas pc$^{-2}$ and above (roughly equivalent to extinctions of A$_{\rm V} \sim 7$ or A$_{\rm K} \sim 0.8$).

\item We show that for turbulent clouds, the observed, line-of-sight extinction is different from the sky-averaged (or ``effective'') extinction seen by the gas. For an observed extinction of around A$_{\rm V} = 7$, we find that the effective extinction is around A$_{\rm V, eff} \sim 1$--2.
\end{itemize}

Finally, we stress that in this study, we have restricted our attention to clouds in an environment similar to that in the local ISM. The question of whether there is a higher column density threshold for star formation in clouds that are located in more extreme environments, such as the Central Molecular Zone of the Milky Way \citep{long13}, is an interesting one, but lies outside of the scope of our present effort.

\section*{Acknowledgements}
We would like to thank the referee, Chris McKee, for an insightful report that helped us to greatly improve the paper. The authors would also like to thank Charles Lada, Mark Krumholz, Mordecai-Mark {Mac Low}, Eve Ostriker, Rahul Shetty, Ralf Klessen, Bruce Elmegreen and Tom Megeath for many enlightening discussions about our results. We would also like to thank the participants of the Heidelberg meeting on ``Galactic Scale Star Formation'' (30th July to 3rd August 2012) for providing the stimulating environment that motivated this paper. The authors acknowledge financial support from the Deutsche Forschungsgemeinschaft (DFG) via SFB 881 ``The Milky Way System'' (sub-projects B1, B2 and B8), and from the Baden-W\"{u}rttemberg-Stiftung by contract research via the programme Internationale Spitzenforschung II (grant P-LS-SPII/18). PCC is supported by grant CL 463/2-1, part of the DFG priority program 1573 ``Physics of the Interstellar Medium''. The simulations described in this paper were performed on the {\em kolob} cluster at the University of Heidelberg, which is funded in part by the DFG via Emmy-Noether grant BA 3706, and via a Frontier grant of Heidelberg University, sponsored by the German Excellence Initiative as well as the Baden-W\"urttemberg Foundation.


\begin{thebibliography}{}

\bibitem[Aguti et~al.(2007)]{aguti07}
Aguti, E.~D., Lada, C.~J., Bergin, E.~A., Alves, J.~F., \& Birkinshaw, M.\ 2007, ApJ, 665, 457

\bibitem[Alves et 
al.(2007)]{alves2007} Alves, J., Lombardi, M., \& Lada, C.~J.\ 2007, A\&A, 462, L17 

\bibitem[Banerjee et~al.(2009)]{banerjee09}
Banerjee, R., V\'{a}zquez-Semadeni, E., Hennebelle, P., \& Klessen, R.~S. 2009, MNRAS, 398, 1082

\bibitem[Bate, Bonnell \& Price(1995)]{bbp95}
Bate, M.~R., Bonnell, I.~A., \& Price, N.~M.\ 1995, MNRAS, 277, 362

\bibitem[Bate \& Burkert(1997)]{bb97} Bate, M.~R., \& Burkert, A.\ 1997, MNRAS, 288, 1060 

\bibitem[Bate(2012)]{Bate2012} Bate, M.~R.\ 2012, MNRAS, 419, 
3115 

\bibitem[Benz(1990)]{benz90}
Benz, W., 1990, in `Proceedings of the NATO Advanced Research Workshop on The
Numerical Modelling of Nonlinear Stellar Pulsations Problems and Prospects',
ed.\ J.~R.~Buchler, (Dordrecht: Kluwer), 269

\bibitem[Bergin et~al.(2001)]{bergin01}
Bergin, E.~A., Ciardi, D.~R., Lada, C.~J., Alves, J., \& Lada, E.~A.\ 2001, ApJ, 557, 209

\bibitem[Bergin et~al.(2002)]{bergin02}
Bergin, E.~A., Alves, J., Huard, T., \& Lada, C.~J.\ 2002, ApJ, 570, L101

\bibitem[Bethell, Zweibel \& Li(2007)]{bethell07}
Bethell, T.~J., Zweibel, E.~G., \& Li, P.~S.\ 2007, ApJ, 667, 275

\bibitem[Bigiel et al.(2008)]{big08} 
Bigiel, F., Leroy, A., Walter, F., Brinks, E., de Blok, W.~J.~G., Madore, B., 
\& Thornley, M.~D.\ 2008, AJ, 136, 2846

\bibitem[Bigiel et al.(2011)]{big11}
Bigiel, F., Leroy, A.~K., Walter, F., Brinks, E., {de Blok}, W.~J.~G., Kramer, C., Rix, H.~W., 
Schruba, A., Schuster, K.~F., Usero, A., \& Wiesemeyer, H.~W.\ 2011, ApJ, 730, L13

\bibitem[Black(1994)]{bl94}
Black, J.~H. 1994, ASP Conf.\ Ser.\ 58, in The First Symposium on the Infrared Cirrus
and Diffuse Interstellar Clouds, eds.\ R.~M.~Cutri \& W.~B.~Latter, (San Francisco:ASP), 355

\bibitem[Clark et al.(2005)]{clark05}
Clark, P.~C., Bonnell, I.~A., Zinnecker, H., \& Bate, M.\ 2005, MNRAS, 359, 809

\bibitem[Clark et~al.(2011)]{clark11}
Clark, P.~C., Glover, S.~C.~O., Klessen, R.~S., \& Bromm, V.\ 2011, ApJ, 727, 110

\bibitem[Clark, Glover \& Klessen(2012)]{cgk12}
Clark, P.~C., Glover, S.~C.~O., \& Klessen, R.~S.\ 2012, MNRAS, 420, 745

\bibitem[Clark et al.(2012)]{clark12}
Clark, P.~C., Glover, S.~C.~O., Klessen, R.~S., \& Bonnell, I.~A.\ 2012, MNRAS, 424, 2599

\bibitem[Collins et~al.(2011)]{collins11}
Collins, D.~C., Padoan, P., Norman, M.~L., \& Xu, H.\ 2011, ApJ, 731, 59

\bibitem[Cunningham et al.(2011)]{Cunningham2011} Cunningham, A.~J., 
Klein, R.~I., Krumholz, M.~R., \& McKee, C.~F.\ 2011, ApJ, 740, 107 

\bibitem[Dale et al.(2012)]{Dale2012} Dale, J.~E., Ercolano, B., 
\& Bonnell, I.~A.\ 2012, MNRAS, 424, 377 

\bibitem[Dale et al.(2013)]{Dale2013} Dale, J.~E., Ngoumou, J., 
Ercolano, B., \& Bonnell, I.~A.\ 2013, MNRAS, 436, 3430 

\bibitem[Dobbs et~al.(2008)]{dobbs08}
Dobbs, C. L., Glover, S.~C.~O., Clark, P.~C., \& Klessen, R.~S.\ 2008, MNRAS, 389, 1097

\bibitem[Dobbs et al.(2011)]{dobbs11} 
Dobbs, C.~L., Burkert, A., \& Pringle, J.~E.\ 2011, MNRAS, 413, 2935

\bibitem[Draine(1978)]{dr78}
Draine, B.~T. 1978, ApJS,  36, 595

\bibitem[Draine \& Bertoldi(1996)]{db96}
Draine, B.~T., \& Bertoldi, F.\ 1996, ApJ, 468, 269

\bibitem[Elmegreen(2000)]{elmegreen00}
Elmegreen, B.~G.\ 2000, ApJ, 530, 277

\bibitem[Evans et al.(2009)]{evans09}
Evans, N.~J., et~al.\ 2009, ApJS, 181, 321

\bibitem[Federrath \& Klessen(2012)]{fk12}
Federrath, C., \& Klessen, R.~S.\ 2012, ApJ, 761, 156

\bibitem[Federrath \& Klessen(2013)]{fk13} 
Federrath, C., \& Klessen, R.~S.\ 2013, ApJ, 763, 51

\bibitem[Federrath et~al.(2008)]{fed08}
Federrath, C., Glover, S.~C.~O., Klessen, R.~S., \& Schmidt, W.\ 2008, Phys.\ Scripta, 132, 04025

\bibitem[Federrath, Klessen, \& Schmidt(2008)]{fks08}
Federrath, C., Klessen, R.~S., \& Schmidt, W.\ 2008, ApJ, 688, L79


\bibitem[Gao \& Solomon(2004a)]{gs04a} Gao, Y., \& Solomon, P.~M.\ 2004, ApJS, 152, 63 


\bibitem[Gao \& Solomon(2004b)]{gs04b} Gao, Y., \& Solomon, P.~M.\ 2004, ApJ, 606, 271 

\bibitem[Glover \& Clark(2012a)]{gc12a}
Glover, S.~C.~O., \& Clark, P.~C.\ 2012a, MNRAS, 421, 9

\bibitem[Glover \& Clark(2012b)]{gc12b}
Glover, S.~C.~O., \& Clark, P.~C.\ 2012b, MNRAS, 421, 116

\bibitem[Glover \& Clark(2012c)]{gc12c}
Glover, S.~C.~O., \& Clark, P.~C.\ 2012c, MNRAS, 426, 377

\bibitem[Glover et al.(2010)]{g10}
Glover, S.~C.~O., Federrath, C., {Mac Low}, M.-M., \& Klessen, R.~S.\ 2010, MNRAS, 404, 2

\bibitem[Glover \& {Mac Low}(2007a)]{gm07a}
Glover, S.~C.~O., \& {Mac Low}, M.-M.\ 2007, ApJS, 169, 239

\bibitem[Glover \& {Mac Low}(2007b)]{gm07b}
Glover, S.~C.~O., \& {Mac Low}, M.-M.\ 2007, ApJ, 659, 1317

\bibitem[Goldsmith(2001)]{gold01}
Goldsmith, P.~F. 2001, ApJ, 557, 736

\bibitem[Gutermuth et~al.(2011)]{guter11}
Gutermuth, R.~A., Pipher, J.~L., Megeath, S.~T., Myers, P.~C., Allen, L.~E., \& Allen, T.~S.\ 2011, ApJ, 739, 84

\bibitem[Heiderman et al.(2010)]{heid10}
Heiderman, A., Evans, N.~J., Allen, L.~E., Huard, T., \& Heyer, M.\ 2010, ApJ, 723, 1019

\bibitem[Heitsch et al.(2006)]{Heitsch2006} 
Heitsch, F., Slyz, A.~D., Devriendt, J.~E.~G., Hartmann, L.~W., 
\& Burkert, A.\ 2006, ApJ, 648, 1052 

\bibitem[Hennebelle et al.(2008)]{Hennebelle2008} 
Hennebelle, P., Banerjee, R., V{\'a}zquez-Semadeni, E., Klessen, R.~S., \& Audit, E.\ 2008, A\&A, 486, L43

\bibitem[Hennebelle 
\& Chabrier(2011)]{HC2011} Hennebelle, P., \& Chabrier, G.\ 2011, ApJ, 743, L29


\bibitem[Heyer et~al.(2009)]{heyer09}
Heyer, M., Krawczyk, C., Duval, J., \& Jackson, J.~M.\ 2009, ApJ, 699, 1092

\bibitem[Hollenbach \& McKee(1979)]{hm79} 
Hollenbach, D., \& McKee, C.~F.\ 1979, ApJS, 41, 555

\bibitem[Indriolo \& McCall(2012)]{ind12}
Indriolo, N., \& McCall, B.~J.\ 2012, ApJ, 745, 91

\bibitem[Jappsen et~al.(2005)]{jappsen05}
Jappsen, A.-K., Klessen, R.~S., Larson, R.~B., Li, Y., {Mac Low}, M.-M.\ 2005,
A\&A, 435, 611

\bibitem[Kennicutt(1998)]{k98}
Kennicutt, R.~C.\ 1998, ARA\&A, 36, 189

\bibitem[Kennicutt et al.(2007)]{ken07}
Kennicutt, R.~C., et~al.\ 2007, ApJ, 671, 333

\bibitem[Kennicutt \& Evans(2012)]{ke12}
Kennicutt, R.~C., Evans, N.~J.\ 2012, ARA\&A, 50, 531

\bibitem[Klessen(2000)]{klessen00}
Klessen, R.~S., 2000, ApJ, 535, 869

\bibitem[Klessen, Heitsch \& {Mac Low}(2000)]{khm00}
Klessen, R.~S., Heitsch, F., \& {Mac Low}, M.-M.\ 2000,ApJ, 535, 887

\bibitem[Kritsuk, Norman \& Wagner(2011)]{krit11}
Kritsuk, A.~G., Norman, M.~L., \& Wagner, R.\ 2011, ApJ, 727, L20

\bibitem[Krumholz, Leroy \& McKee(2011)]{klm11}
Krumholz, M.~R., Leroy, A.~K., \& McKee, C.~F.\ 2011, ApJ, 731, 25

\bibitem[Krumholz \& McKee(2005)]{km05}
Krumholz, M.~R., McKee, C.~F.\ 2005, ApJ, 630, 250

\bibitem[Krumholz \& Tan(2007)]{kt07}
Krumholz, M.~R., \& Tan, J.~C.\ 2007, ApJ, 654, 304

\bibitem[Krumholz et al.(2012)]{kdm2012} 
Krumholz, M.~R., Dekel, A., \& McKee, C.~F.\ 2012, ApJ, 745, 69 

\bibitem[Krumholz et al.(2012)]{kkm2012} Krumholz, M.~R., 
Klein, R.~I., \& McKee, C.~F.\ 2012, ApJ, 754, 71 

\bibitem[Lada, Lombardi \& Alves(2010)]{lada10}
Lada, C.~J., Lombardi, M., \& Alves, J.~F.\ 2010, ApJ, 724, 687

\bibitem[Lada et al.(2012)]{lada12}
Lada, C.~J., Forbrich, J., Lombardi, M., \& Alves, J.~F.\ 2012, ApJ, 745, 190

\bibitem[Larson(1981)]{larson81}
Larson, R.~B.\ 1981, MNRAS, 194, 809

\bibitem[Lee et~al.(1996)]{lee96}
Lee, H.-H., Herbst, E., {Pineau des For\^ets}, G.,
Roueff, E., \& {Le Bourlot}, J.\ 1996, A\&A, 311, 690

\bibitem[Liu et al.(2011)]{liu11}
Liu, G., Koda, J., Calzetti, D., Fukuhara, M., \& Momose, R.\ 2011, ApJ, 735, 63

\bibitem[Longmore et~al.(2013)]{long13}
Longmore, S.~N., et~al.\ 2013, MNRAS, 429, 987

\bibitem[Machida et al.(2009)]{Machida2009} Machida, M.~N., 
Inutsuka, S.-i., \& Matsumoto, T.\ 2009, ApJ, 699, L157

\bibitem[Machida 
\& Matsumoto(2012)]{Machida2012} Machida, M.~N., \& Matsumoto, T.\ 2012, MNRAS, 421, 588

\bibitem[{Mac Low}(1999)]{maclow99}
{Mac Low}, M.-M.\ 1999, ApJ, 524, 169

\bibitem[{Mac Low} \& Klessen(2004)]{mlk2004} 
Mac Low, M.-M., \& Klessen, R.~S.\ 2004, Reviews of Modern Physics, 76, 125 

\bibitem[Maret \& Bergin(2007)]{mb07}
Maret, S., \& Bergin, E.~A.\ 2007, ApJ, 664, 956

\bibitem[Matzner 
\& McKee(2000)]{mm2000} Matzner, C.~D., \& McKee, C.~F.\ 2000, ApJ, 545, 364

\bibitem[McCall et~al.(2003)]{mac03}
McCall, B.~J., et~al.\ 2003, Nature, 422, 500

\bibitem[McKee \& Offner(2010)]{McKeeOffner2010} McKee, C.~F., \& Offner, S.~S.~R.\ 2010, ApJ, 716, 167

\bibitem[Molina et~al.(2011)]{molina11}
Molina, F., Glover, S., \& Federrath, C.\ 2011, in ``Conditions and Impact of Star Formation: 
New Results with Herschel and Beyond'',  EAS Publications Series 52, 
eds.\ M.~R{\"o}llig, R.~Simon, V.~Ossenkopf, \& J.~Stutzki, 289

\bibitem[Molinari et~al.(2014)]{molinari14}
Molinari, S., Bally, J., Glover, S., Moore, T., Noriega-Crespo, A., Plume, R., Testi, L., 
V\'azquez-Semadeni, E., Zavagno, A., Bernard, J.-P., \& Martin, P., 2014, to appear
in ``Protostars and Planets VI'', eds.\ H.~Beuther, R.~S.~Klessen, C.~P.~Dullemond (University
of Arizona Press); arXiv:1402.6196

\bibitem[Narayanan et al.(2012)]{nara12}
Narayanan, D., Krumholz, M.~R., Ostriker, E.~C., \& Hernquist, L.\ 2012, MNRAS, 421, 3127

\bibitem[Nelson \& Langer(1999)]{nl99}
Nelson, R.~P., \& Langer, W.~D. 1999, ApJ, 524, 923

\bibitem[Offner 
\& McKee(2011)]{offner11} Offner, S.~S.~R., \& McKee, C.~F.\ 2011, ApJ, 736, 53 

\bibitem[Offner et al.(2012)]{offner12} Offner, S.~S.~R., 
Robitaille, T.~P., Hansen, C.~E., McKee, C.~F., 
\& Klein, R.~I.\ 2012, ApJ, 753, 98 

\bibitem[Padoan et al.(2012)]{phn2012} Padoan, P., 
Haugb{\o}lle, T., \& Nordlund, {\AA}.\ 2012, ApJ, 759, L27 

\bibitem[Padoan \& Nordlund(2011)]{PN11}
Padoan, P., \& Nordlund, \AA.\ 2011, ApJ, 730, 40

\bibitem[Price 
\& Bate(2008)]{PriceBate2008} Price, D.~J., \& Bate, M.~R.\ 2008, MNRAS, 385, 1820

\bibitem[Price et al.(2012)]{Price2012} Price, D.~J., Tricco, 
T.~S., \& Bate, M.~R.\ 2012, MNRAS, 423, L45

\bibitem[Rosas-Guevara et~al.(2010)]{rg10}
Rosas-Guevara, Y., V\'{a}zquez-Semadeni, E., G\'{o}mez, G.~C., Jappsen, A.-K.\ 2010, MNRAS, 406, 1875

\bibitem[Roman-Duval et~al.(2010)]{rd10}
Roman-Duval, J., Jackson, J.~M., Heyer, M., Rathborne, J., \& Simon, R.\ 2010, ApJ, 723, 492

\bibitem[Schaye(2004)]{schaye04} 
Schaye, J.\ 2004, ApJ, 609, 667

\bibitem[Schmidt(1959)]{schmidt59}
Schmidt, M. 1959, ApJ, 129, 243 

\bibitem[Schruba et~al.(2011)]{schr11}
Schruba, A., et~al.\ 2011, AJ, 142, 37

\bibitem[Shaw et~al.(2008)]{shaw08}
Shaw, G., Ferland, G.~J., Srianand, R., Abel, N.~P., {van Hoof}, P.~A.~M., \& Stancil, P.~C.\ 2008, ApJ, 675, 405

\bibitem[Shetty et al.(2013)]{shetty13a}
Shetty, R., Kelly, B.~C., \& Bigiel, F.\ 2013, MNRAS, 430, 288

\bibitem[Shetty et al.(2014)]{shetty13b}
Shetty, R., Kelly, B.~C., Rahman, N., Bigiel, F., Bolatto, A.~D., Clark, P.~C., Klessen, R.~S., \&
Konstandin, L.~K.\ 2014, MNRAS, 437, L61

\bibitem[Shu(1977)]{Shu1977} Shu, F.~H.\ 1977, ApJ, 214, 488

\bibitem[Slyz et~al.(2005)]{slyz05}
Slyz, A.~D., Devriendt, J.~E.~G., Bryan, G., \& Silk, J.\ 2005, MNRAS, 356, 737

\bibitem[Solomon et al.(1987)]{Solomon1987} 
Solomon, P.~M., Rivolo, A.~R., Barrett, J., \& Yahil, A.\ 1987, ApJ, 319, 730

\bibitem[Springel(2005)]{springel05}
Springel, V.\ 2005, MNRAS, 364, 1105

\bibitem[Tan et al.(2006)]{Tan2006} Tan, J.~C., Krumholz, 
M.~R., \& McKee, C.~F.\ 2006, ApJL, 641, L121 

\bibitem[{van der Tak} \& {van Dishoeck}(2000)]{vdt00}
{van der Tak}, F.~F.~S., \& {van Dishoeck}, E.~F.\ 2000, A\&A, 358, L79

\bibitem[V{\'a}zquez-Semadeni et al.(2006)]{enrique2006} 
V{\'a}zquez-Semadeni, E., Ryu, D., Passot, T., Gonz{\'a}lez, R.~F., 
\& Gazol, A.\ 2006, ApJ, 643, 245 

\bibitem[V\'{a}zquez-Semadeni et al.(2007)]{vs07}
V\'{a}zquez-Semadeni, E., G\'omez, G., Jappsen, A.-K., Ballesteros-Paredes, J., 
Gonz\'{a}lez, R.~F., \& Klessen, R.~S.\ 2007, ApJ, 657, 870

\bibitem[V{\'a}zquez-Semadeni et al.(2008)]{vs08}
V{\'a}zquez-Semadeni, E., Gonz\'{a}lez, R.~F., Ballesteros-Paredes, J., 
Gazol, A., \& Kim, J.\ 2008, MNRAS, 390, 769

\bibitem[V{\'a}zquez-Semadeni et al.(2009)]{enrique2009} 
V{\'a}zquez-Semadeni, E., G{\'o}mez, G.~C., Jappsen, A.-K., 
Ballesteros-Paredes, J., \& Klessen, R.~S.\ 2009, ApJ, 707, 1023 

\bibitem[Wang et al.(2010)]{wang2010} Wang, P., Li, Z.-Y., Abel, 
T., \& Nakamura, F.\ 2010, ApJ, 709, 27

\bibitem[Wolfire et al.(1995)]{wolf95}
Wolfire, M.~G., Hollenbach, D., McKee, C.~F., Tielens, A.~G.~G.~M., \& Bakes, E.~L.~O.\ 1995,
ApJ, 443, 152

\bibitem[Wu et al.(2005)]{wu05} Wu, J., Evans, N.~J., II, 
Gao, Y., et al.\ 2005, ApJL, 635, L173

\end{thebibliography}
\end{document}